\documentclass[10pt, final, journal, letterpaper, twocolumn]{IEEEtran}

\usepackage[dvips]{graphicx}
\usepackage{times}
\usepackage{cite}
\usepackage{amsmath}
\usepackage{array}
\usepackage{amssymb}

\usepackage{stfloats}
\usepackage{rotating,threeparttable,booktabs}
\usepackage{bm}
\usepackage{dcolumn,booktabs}
\usepackage{multirow}
\usepackage{graphicx}
\usepackage{subfigure}
\usepackage{color}

\usepackage{amsmath,amsthm,amssymb,amsfonts}
\makeatletter
\thm@headfont{\sc}
\makeatother

\usepackage{url}
\begin{document}
\title{An Overview on Multiway Relay Communications: Fundamental Issues, Recent Advances, and New Challenges}
\author{\authorblockA {Rui Wang, \textit{Senior Member IEEE} and Xiaojun Yuan, \textit{Senior Member IEEE}}
\thanks{R. Wang is with the Department of Information and Communications at Tongji University, Shanghai, 201804, P. R. China (e-mail: ruiwang@tongji.edu.cn). }
\thanks{X. Yuan is with the Center for Intelligent Networking and Communications, the National Key Laboratory of Science and Technology on Communications, and the University of Electronic Science and Technology of China, Chengdu, China (e-mail: xjyuan@uestc.edu.cn).}}
\maketitle

\begin{abstract}
To address the coverage and capacity challenges in fifth-generation (5G) wireless networks, the relay technique has been considered as a promising solution and recently attracted great attentions from academia. The multiway relay channel (mRC), which serves as a fundamental building block of a relay network, defines the family of all possible information-flow patterns between a relay node and its neighboring nodes. Combined with several new techniques, such as physical-layer network coding (PLNC), signal and interference alignments, multi-input multi-output (MIMO), etc., the mRC exhibits superior capability to assist multiuser information exchange. In this article, we aim to provide a comprehensive overview on the state-of-the-art of the mRC from the perspectives of both information-theoretic and communication-theoretic studies. The key results are summarized and discussed to identify the current research progress of the mRC. Open problems and research challenges are also discussed to spur future research towards the mRC.
\end{abstract}


\section{Introduction}

The commercialization of fifth-generation (5G) wireless communications has been launched since 2019. The vision of 5G is to provide services of multigigabit-per-second data rates with ultra-low latency and massive terminal connections. 
To achieve this goal, a much broader spectrum, especially frequency bands beyond 6 GHz, needs to be put in use to support 5G. For example, the first phase of the standardization of the 5G radio access technology by the 3rd Generation Partnership Project (3GPP) has confirmed to support carrier frequencies up to 52.6 GHz \cite{Polese_2020_COMM}\footnote{The mmWave band is also expected to support sixth-generation (6G) in the near future \cite{Saad_IEEENet2019}.}. However, wireless communications operating at millimetre-wave (mmWave) bands comes with several unprecedented challenges including severe path and penetration losses, and so line-of-sight (LoS) propagation is usually dominant in mmWave channels.
LOS may easily be blocked by moving obstacles such as human beings, vehicles, and even falling leaves due to short wavelength of mmWaves.
As such, mmWave radio access networks suffer from a severe signal coverage problem as compared with sub-6GHz wireless networks.

To address the above challenge, the relay technique has been considered as a promising solution to vastly boost 5G mmWave radio network coverage and capacity with low energy consumption and networking costs. In specific, the relay technique has been standardized by the 3GPP to support integrated access and backhaul (IAB) as a cost-effective alternative to wired backhaul. With IAB, a large portion of next-generation base stations (BSs) will be able to wirelessly relay the backhaul traffic through multiple hops at mmWave frequencies, which greatly reduces the deployment cost as compared to other candidate techniques such as using high directional gain antennas and using network densification \cite{Polese_2020_COMM}. Moreover, relay cooperative transmission has also been approved by 3GPP RAN$\#$86 to support sidelink transmission to address automotive industry and critical communication demands \cite{3GPP_R17_project}.

The multiway relay channel (mRC), which serves as a fundamental building block of a relay network, generally defines the family of all possible information-flow topologies between a relay node and its neighbouring nodes (usually referred to as user nodes). The mRC faces a couple of unique challenges. First, a relay node typically operates in a half-duplex mode, i.e., a relay node receives and transmits signals in orthogonal time or frequency resource blocks, which causes the sacrifice of spectrum efficiency. Second, multiple user nodes in the mRC simultaneously send information to a common relay, which causes multiuser interference. To address these challenges, abundant new techniques, such as physical-layer network coding (PLNC), signal and interference alignment, multi-input multi-output (MIMO), etc., have been developed in the literature in the past decade. Here we briefly introduce these techniques as follows.

PLNC is an efficient technique to combat the spectrum sacrifice problem of the mRC by fundamentally changing the communication paradigm. PLNC, originally proposed in \cite{PLNC_1_ZhangACM2006}, can be considered as a physical layer application of the idea of network coding \cite{PLNC_11_Ahlswede}. Exploiting the fact that the electromagnetic (EM) waves are superimposed in the physical space, PLNC allows the relay to decode and retransmit network-coded signals in the physical layer, and the users to decode desired messages by exploiting the knowledge of self-interference. This breaks the traditional belief that interference is regarded as a destructive phenomenon \cite{PLNC_2_Liew2013}, and significantly reduces the time/frequency resource required for completing information exchange, as compared to traditional relay assisted communications. Particularly, as a special case of the mRC, the two-way relay channel (TWRC) models a two-user information exchange under the assistance of the relay. The PLNC technique doubles the spectrum efficiency of the TWRC by reducing the required time slots from four to two in each round of information exchange.

The MIMO technique is widely used to achieve multiplexing in wireless communications. For a MIMO mRC, the deployment of antenna arrays allows the relay and the users to manipulate their desired signals and interference with spatial degrees of freedom. This gives rise to the idea of signal and interference alignment \cite{DoF_Linear_27_LeeTIT2010, DoF_Linear_3_LiuICC2015, DoF_Linear_11_LeeTWC2012,DoF_Linear_31_LiuTIT2015, DoF_Linear_34_WangTSP2014}, \cite{DoF_Linear_15_WangISIT2014, DoF_Linear_2_WangeUC2013}: A relay/user node aims to align its desired signals in certain spatial directions to facilitate network decoding; and at the same time, it aims to align undesired signals in certain spatial directions for interference suppression.

Despite the extensive efforts made towards the understanding of the mRC, a comprehensive overview of exiting studies on the mRC is still lacking.
In this article, we aim to focus on this issue.
By reviewing the existing works, we summarize the key results of the state-of-the-art studies and identify the current research progress from the aspects of information-theoretic and communication-theoretic studies. Furthermore, we discuss some important research directions and challenges.

The remainder of the article is organized as follows. In Section \ref{application_scenario}, the main application scenarios of the mRC are described. The definition of the mRC is presented in Section \ref{system_model}, which covers widely studied data exchange models as some special cases. In Section \ref{information_theoretic}, we discuss the analytical results based on various performance measures, including achievable rate/capacity, degrees-of-freedom, bit-error rate, outage probability and diversity-and-multiplexing trade-off. In Section \ref{communication_theoretic}, we discuss the mRC from the perspective of physical-layer communication techniques, such as channel estimation, power allocation, relay selection, and precoding/beamforming. In Section VI, we discuss open research directions including synchronization, channel state information (CSI) acquisition, channel-unaware mRC, multi-relay/multi-hop mRC, applications in mobile edge computing (MEC), fog radio access network (Fog-RAN) and intelligent reflecting surface (IRS). Finally, the article is concluded in Section \ref{conclusion}.

\section{Application Scenarios}\label{application_scenario}
In this section, several typical application scenarios of the mRC are described. A common feature of these applications is that the information exchange between user terminals is assisted by one or multiple relays. PLNC is employed to improve the spectrum efficiency of the system.

\subsection{Cellular Communications}

\begin{figure}
        \centering
        \includegraphics[width=3.0in]{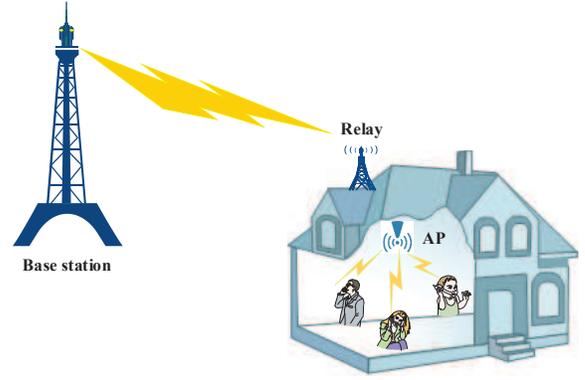}\\
        \caption{The mRC model for cellular communications}
        \label{Application_cellular}
\end{figure}

With the growing demand of mobile data services and internet of things (IoT),  wireless cellular communications is now undergoing a deep transition and faces new challenges, such as higher throughput, larger wireless coverage, and lower power consumption. To address these issues, relay is a promising technology that improves the robustness of the network and supports the access of massive terminals. The advantage of the relay technique has been recognized by the 3GPP in LTE-A standard Release 10 \cite{3GPP_Relay}.

Millimeter-wave frequency bands (beyond 6 GHz) will be used to meet the high throughput demands of next generation wireless services \cite{Rappaport2013_access, Zhao2014_ICM, Zhang_Vasilakos_Hanzo2015}. However, due to the high path degradation and low penetration capability of the electromagnetic wave in high frequency bands, the physical distance to support reliable wireless communications is greatly shortened, which significantly reduces the coverage capability of wireless services. The use of relay provides an efficient and economic solution to address this coverage problem.

A typical scenario is illustrated in Fig. \ref{Application_cellular}, where users in a building intend to communicate with a BS outside in the millimeter-wave frequency band. To compensate the path degradation, a relay is deployed on the building roof, connecting wireless access point (AP) in the building via a wire line. The users, the wireless AP, and the base station form a multiway relay channel.

\subsection{Satellite Communications}

\begin{figure}
        \centering
        \includegraphics[width=3.5 in]{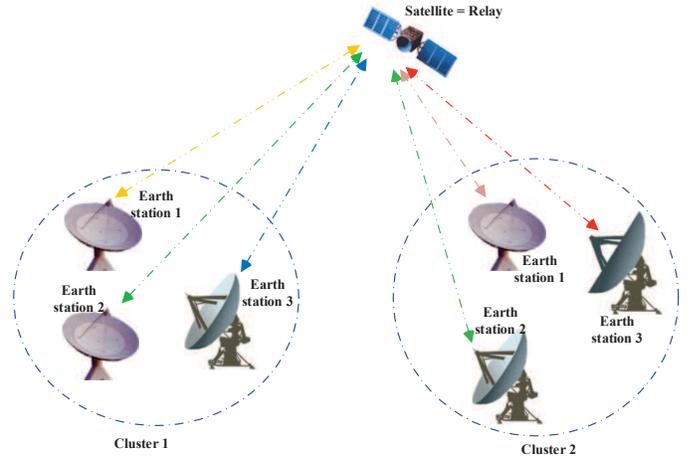}\\
        \caption{The mRC model for satellite communications.}
        \label{Application_Satellite}
\end{figure}

Satellite communications is an important wireless technology due to the special benefits like wide frequency band, large coverage area and line-of-sight (LOS) transmission. As a result, satellite communication has important applications in navigation assistance and disaster recovery. Particularly in disaster recovery, when the terrestrial infrastructure is destroyed, the geostationary earth orbit satellites is able to provide broadband communication links in the disaster area to establish broadband access from the disaster area to the rest of the world.

The mRC can be naturally applied to satellite communications where the satellite, serving as a relay node situated far from the earth in the free space, receives signals from source earth stations and forwards them to destination earth stations. As reported in \cite{Arti_CL2014, Bhatnagar_TCOM2015, Miridakis_TVT2015, Baofeng_Chinacom2011}, two-way relaying has been used to assist the data exchange of two earth terminals. 
Fig. \ref{Application_Satellite} illustrates an example with multiple earth stations exchanging information. In specific, the source stations are classified into two clusters, and only
the source stations in a common cluster intend to exchange data with each other by following an arbitrary traffic pattern.
In general, each cluster may span a large terrestrial distance. The inter-cluster interference can be eliminated, e.g., by using multi-antenna beamforming techniques.

\subsection{Underwater Communications}

\begin{figure}
        \centering
        \includegraphics[width=3.5 in]{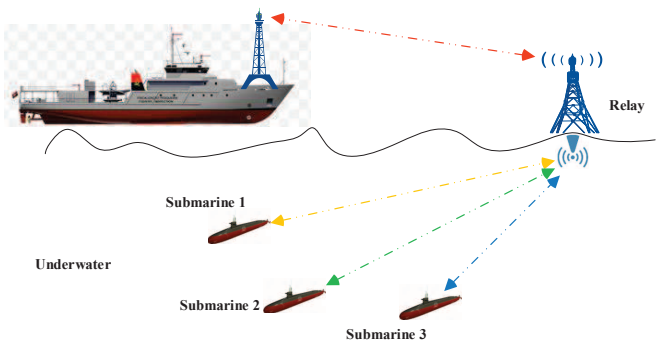}\\
        \caption{The mRC model for underwater communications.}
        \label{Application_Underwater}
\end{figure}

Special tasks such as oceanographic data collection, pollution and environmental monitoring, offshore exploration and disaster prevention, requires to deploy underwater equipments and robot in the deep water to perform tasks that cannot be handled by human. Also, for military applications, submarines always need to stay in the deep ocean to complete missions. To control, localize, and track these underwater equipments, or assign missions to the soldiers in a submarines, wireless communication techniques which work reliably in both air and water need to be developed. Using electromagnetic waves for underwater communications is not feasible due to the high attenuation and scattering effect of the underwater wireless channel. In fact, radio frequency (RF) waves can propagate through salty water only at extremely low frequencies (usually 30-300Hz), which requires large-size antennas and high transmission power. To support long distance underwater communications, the acoustic wireless technique is regarded as a more effective solution \cite{Pompili_COMG2009}.

To realize communications between abovewater and underwater devices, a hybrid wireless technology is required to achieve the reliable communications across both the air and underwater environments. That is, the radio technique is used for wireless communications over the air, while the acoustic technique is used for wireless communications under the water. An intermediate relay node is used to serve as a bridge across two prorogation media. Fig. \ref{Application_Underwater} illustrates a typical scenario for submarine communications, where the mRC can be used to describe this relay-assisted abovewater and underwater communication model.

\subsection{Maritime Communications}

\begin{figure}
 	\centering
 	\includegraphics[width= 3 in]{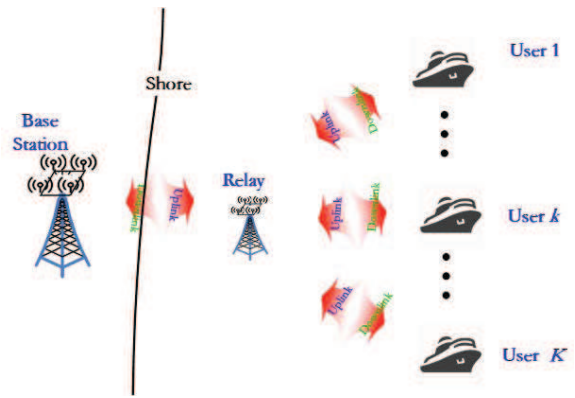}  \vspace{-2mm}
 	\caption{{The mRC model for maritime communications.}}
    \label{Application_Maritime}
 \end{figure}

 \begin{figure}
 	\centering
 	\includegraphics[width= 3.5 in]{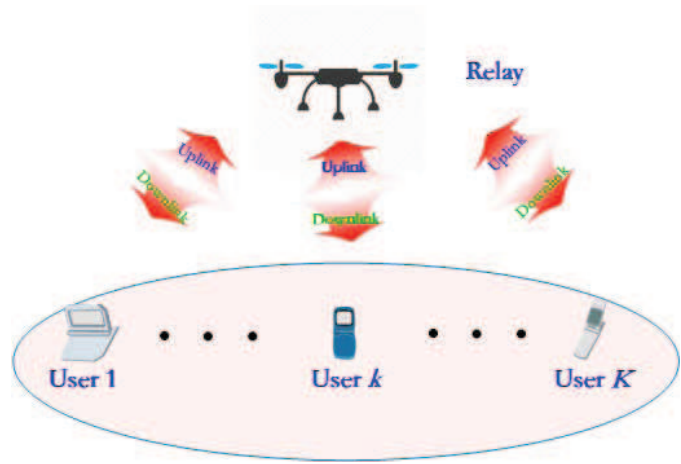}  \vspace{-2mm}
	\caption{{The mRC model for UAV communications.}}
    \label{Application_UAV}
\end{figure}

With the development of maritime economy, there are more and more human activities in the ocean such as environment monitoring, fishery, tactical surveillance, cruise, offshore exploration, etc.  Existing maritime technologies mostly rely on both satellite communications and very-high-frequency (VHF) communications \cite{Bekkadal2009}. However, the corresponding communication cost is extremely high, and the maritime communication system depending on VHT communications cannot provide mobile multimedia services due to its limited bandwidth. Therefore, more attentions have been focused on the exploration of applying terrestrial communication techniques to maritime communications in order to supply low-cost and high data-rate mobile services \cite{Duan2018}\cite{Xu2017}.
Due to their ability of extending wireless network coverage area and improving spatial cooperation diversity, relay communications,
a widely used terrestrial communication technique, has been migrated to the ocean scenario for supporting efficient and reliable information services.
Fig. \ref{Application_Maritime} depicts a typical maritime scenario where one shore-based base station communicates with $K$ maritime users via the help of an offshore relay, as an example of the mRC.


\subsection{Unmanned Aerial Vehicle (UAV) Communications}

Nowadays, UAV has been used frequently in a large variety of applications \cite{Union2009}.
As UAVs usually fly at high altitudes, the on-the-fly communications established by UAVs are line-of-sight (LoS) links, which help mitigate signal blockage and signal shadowing.
Therefore, more and more attention has been focused on the use of UAVs as flying wireless communication platforms \cite{Mozaffari2017, Gupta2016}.
In particular, UAVs can be used as wireless relays to form a mRC to enhance the connectivity and extend coverage of ground wireless services.
Fig. \ref{Application_UAV} illustrates
a typical UAV-aided mRC communication scenario, where a single UAV serves as a relay to assist data exchange between $K$ ground devices.

\begin{figure*}
        \centering
        \includegraphics[width=4.0in]{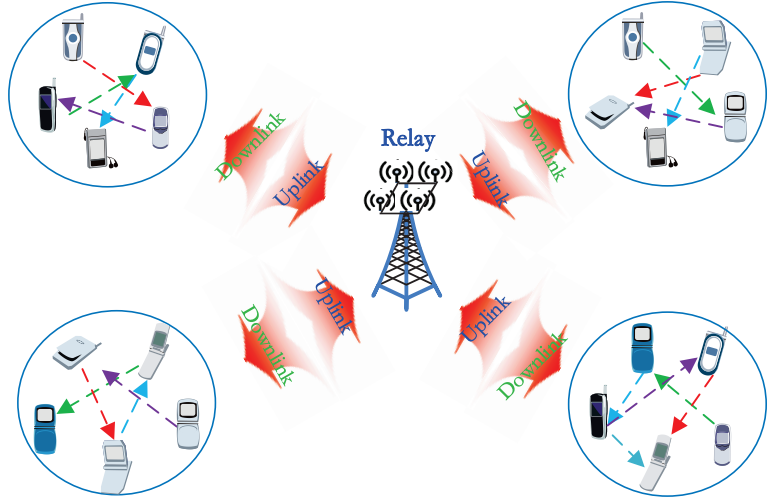}\\
        \caption{An illustration of the mRC.}
        \label{System Model}
\end{figure*}

\begin{figure*}
        \centering
        \includegraphics[width=6.5in]{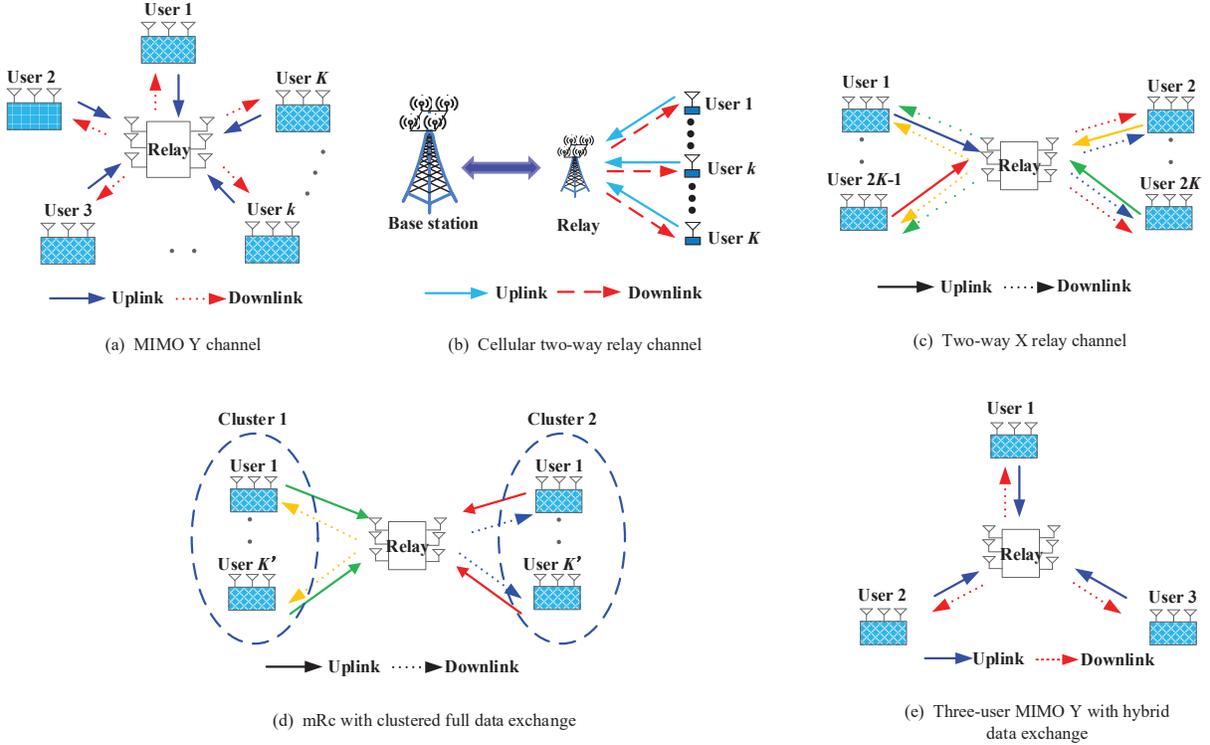}\\
        \caption{Some special cases of the mRC}.
        \label{Special_case}
\end{figure*}

\section{Multiway Relay Model}\label{system_model}
In this section, we show how to mathematically describe a general data exchange model for multiway relay communications. We also discuss how to specialize the general data exchange model to the ones studied in the literature and present some typical relay strategies used for the mRC.


\subsection{General Channel Model}

Without loss of generality, it is assumed that the considered MIMO mRC includes $K$ users for information exchange. Due to deep channel fading and shadowing effect, the direct links between users are not available. The data exchange among $K$ users is assisted by a common relay node.
Each user $k$ has $M_k$ antennas and the relay has $N$ antennas with $M_k\geq 1$ and $N\geq 1$.\footnote{The considered channel model reduces to the single-antenna case by letting $M_1=M_2=\cdots=M_K=N=1$.}  An illustration of the mRC with an arbitrary data exchange model is presented in Fig. \ref{System Model}.

Each information exchange round consists of two phases. The first phase is known as the uplink phase, or alternatively the multiple access (MAC) phase, in which the users simultaneously send signals to the relay. The received signal at the relay is represented as
\begin{equation}\label{System-1}
\mathbf{Y}_R = \sum_{k=1}^{K}\mathbf{H}_{k}\mathbf{X}_{k}+\mathbf{N}_R,
\end{equation}
where $\mathbf{H}_{k}$ denotes the channel matrix between user $k$ and the relay in the uplink phase; $\mathbf{X}_{k}\in \mathcal{\mathbb{C}}^{M_k\times T_{\uparrow}}$ represents the transmit signal from user $k$; $\mathbf{Y}_{R} \in \mathcal{\mathbb{C}}^{N\times T_{\uparrow}}$ is the relay's received signal in the uplink phase; $\mathbf{N}_R \in \mathcal{\mathbb{C}}^{N\times T_{\uparrow}}$ represents the additive white Gaussian noise (AWGN) at the relay; $T_{\uparrow}$ is the time duration of the MAC phase.


The second phase refers to the downlink phase, or alternatively the broadcasting (BC) phase, in which the relay broadcasts signals to the user ends. The $k$th user's received signal is represented as
\begin{equation}\label{System-2}
\mathbf{Y}_{k} = \mathbf{G}^T_{k}\mathbf{X}_{R}+\mathbf{N}_{k},
\end{equation}
where $\mathbf{G}_{k}\in \mathcal{\mathbb{C}}^{M_k\times N}$ represents the channel matrix between the relay and the $k$th user during the BC phase;
$\mathbf{X}_{R}\in \mathcal{\mathbb{C}}^{N\times T_\downarrow}$ denotes the relay's transmit signal during the BC phase; $\mathbf{N}_{k}$ denotes the AWGN matrix at user $k$; $T_\downarrow$ is the time duration of the downlink.

To model an arbitrary data exchange, we use two indicator matrices, i.e., one for uplink and the other for downlink to characterize the data flow in the mRC.
More specifically, we define a $K\times F$ indicator matrix ${\bf D}_{\uparrow}$ to describe the data flow in the uplink:
\begin{equation}\label{System-22}
\begin{split}
 {\bf D}_{\uparrow}  =
 \left(
                                  \begin{array}{ccccc}
                                    D_{\uparrow,1,1}  & \cdots & D_{\uparrow,1,f} & \cdots & D_{\uparrow,1,F} \\
                                    D_{\uparrow,2,1}  & \cdots & D_{\uparrow,2,f} & \cdots & D_{\uparrow,2,F}\\
                                    \vdots &  \vdots & \vdots & \ddots & \vdots\\
                                    D_{\uparrow,K,1}  & \cdots & D_{\uparrow,K,f} & \cdots & D_{\uparrow,K,F}\\
                                  \end{array}
                                \right)
\end{split}
\end{equation}
where $D_{\uparrow,k,f}\in \{0,1\}$ for $\forall k,f$ and $F$ is the total number of messages in the mRC. Note that $D_{\uparrow,k,f}=1$ means that message $f$ is sent from user $k$, and $D_{\uparrow,k,f}=0$ otherwise.

Similarly, we use an $F\times K$ indicator matrix ${\bf D}_{\downarrow}$ to characterize the data flow in the downlink
\begin{equation}\label{System-222}
\begin{split}
 {\bf D}_{\downarrow}  =
 \left(
                                  \begin{array}{cccccc}
                                    D_{\downarrow,1,1}   & \cdots & D_{\downarrow,1,k} & \cdots & D_{\downarrow,1,K} \\
                                    D_{\downarrow,2,1}   & \cdots & D_{\downarrow,2,k} & \cdots & D_{\downarrow,2,K}\\
                                    \vdots & \vdots   & \vdots & \ddots & \vdots\\
                                    D_{\downarrow,F,1}   & \cdots & D_{\downarrow,F,K} & \cdots & D_{\downarrow,F,K}\\
                                  \end{array}
                                \right)
\end{split}
\end{equation}
where $D_{\downarrow,f,k}\in \{0,1\}$ indicates whether message $f$ is sent to user $k$. With the above definition of ${\bf D}_{\uparrow}$ and ${\bf D}_{\downarrow}$, the $(k,k^\prime)$ entry of the product of ${\bf D}_{\uparrow}$ and ${\bf D}_{\downarrow}$ represents the number of independent messages which are transmitted from user $k$ to user $k^\prime$.

\subsection{Use Cases}

In this subsection, we present some use cases of the mRC that have been investigated in the literature.

\subsubsection{MIMO Y channel}
As illustrated in Fig. \ref{Special_case} (a), in the $K$-user MIMO Y channel, each user $k$ exchanges a private message with each of the remaining $K-1$ users.
The MIMO Y channel can be regarded as a simple case of the general mRC described in Subsection A. We take $K=3$ as an example. Denote by $\{s_{1,1}, s_{1,2}, s_{2,1}, s_{2,2}, s_{3,1}, s_{3,2}\}$ the messages exchanged in the channel where $s_{i,j}$ is the $j$th message sent from user $i$.
The uplink and downlink indicator matrices are given by
%
%
%
%
\begin{equation}\label{System-4}
{\bf D}_{\uparrow} =\bordermatrix{%
& s_{1,1} & s_{1,2} & s_{2,1} & s_{2,2} & _{3,1} & s_{3,2} \cr
{\rm user}_1 & 1 & 1 & 0 & 0 & 0 & 0  \cr
{\rm user}_2 & 0 & 0 & 1 & 1 & 0 & 0  \cr
{\rm user}_3 & 0 & 0 & 0 & 0 & 1 & 1  \cr
}\end{equation}
%
and
\begin{equation}\label{System-44}
{\bf D}_{\downarrow} = \bordermatrix{%
& {\rm user}_1 & {\rm user}_2 & {\rm user}_3 \cr
 s_{1,1} &               0 & 1 & 0  \cr
 s_{1,2} &               0 & 0 & 1  \cr
 s_{2,1} &               1 & 0 & 0  \cr
 s_{2,2} &               0 & 0 & 1  \cr
 s_{3,1} &               1 & 0 & 0  \cr
 s_{3,2} &               0 & 1 & 0  \cr
}
\end{equation}
As a result, we have
\begin{equation}\label{System-444}
{\bf D}_{\uparrow} {\bf D}_{\downarrow}  = \left[
              \begin{array}{ccc}
                0 & 1 & 1  \\
                1 & 0 & 1 \\
                1 & 1 & 0 \\
              \end{array}
            \right],
\end{equation}
where each $``1"$ in the $(i,j)$th position indicates that user $i$ sends a message to user $j$.

\subsubsection{Cellular two-way relay channel}
As illustrated in Fig. \ref{Special_case} (b), the cellular TWRC defines transmission protocol where $K$ users exchange private messages with a single base station via a common relay node.
Take $K=2$ as an example and denote by $\{s_{b,1}, s_{b,2}, s_{1}, s_{2}\}$ the messages exchange in the channel where $s_{b,i}$ is the $i$th message sent from base station and $s_{i}$ is the message sent from user $i$.
The uplink and downlink indicator matrices are given by
\begin{equation}\label{System-6}
{\bf D}_{\uparrow} = \left[
              \begin{array}{cccc}
                1 & 1 & 0 & 0   \\
                0 & 0 & 1 &  0   \\
                0 & 0 & 0 & 1   \\
              \end{array}
            \right]
\end{equation}
and
\begin{equation}\label{System-66}
{\bf D}_{\downarrow} = \left[
              \begin{array}{ccc}
                0 & 1 & 0  \\
                0 & 0 & 1  \\
                1 & 0 & 0  \\
                1 & 0 & 0  \\
              \end{array}
            \right],
\end{equation}
with
\begin{equation}\label{System-666}
{\bf D}_{\uparrow} {\bf D}_{\downarrow}  = \left[
              \begin{array}{ccc}
                0 & 1 & 1  \\
                1 & 0 & 0 \\
                1 & 0 & 0 \\
              \end{array}
            \right].
\end{equation}

\subsubsection{Two-way X relay channel}

Fig. \ref{Special_case} (c) depicts a $2K$-user two-way X relay channel with $2K$ users being classified into two groups. Group $1$ includes users $\{1,3,\cdots,2K-1\}$ and group $2$ includes users $\{2,4,\cdots,2K\}$. Each user in a group exchanges a private message with every user in another group.
We take $K=2$ as an example. Denote by $\{s_{1,1}, s_{1,2}, s_{2,1}, s_{2,2}, s_{3,1}, s_{3,2}, s_{4,1}, s_{4,2}\}$ the messages exchange in the channel where $s_{i,j}$ is the $j$th message sent from user $i$.
For $K=2$, the uplink and downlink indicator matrices are given by
\begin{equation}\label{System-7}
{\bf D}_{\uparrow} = \left[
\begin{array}{cccccccc}
                1 & 1 & 0 & 0  & 0 & 0  & 0 & 0  \\
                0 & 0 & 1 &  1 & 0 & 0  & 0 & 0   \\
                0 & 0 & 0 & 0 & 1 & 1  & 0 & 0   \\
                0 & 0 & 0 & 0 & 0 & 0  & 1 & 1   \\
              \end{array}
            \right]
\end{equation}
and
\begin{equation}\label{System-77}
{\bf D}_{\downarrow} = \left[
              \begin{array}{cccc}
                 0 & 1 & 0 & 0 \\
                 0 & 0 & 0 & 1 \\
                 1 & 0 & 0 & 0 \\
                 0 & 0 & 1 & 0 \\
                 0 & 1 & 0 & 0 \\
                 0 & 0 & 0 & 1 \\
                 1 & 0 & 0 & 0 \\
                 0 & 0 & 1 & 0 \\
              \end{array}
            \right],
\end{equation}
with
\begin{equation}\label{System-777}
{\bf D}_{\uparrow} {\bf D}_{\downarrow} = \left[
              \begin{array}{cccc}
                0 & 1 & 0 & 1  \\
                1 & 0 & 1 & 0 \\
                0 & 1 & 0 & 1 \\
                1 & 0 & 1 & 0 \\
              \end{array}
            \right].
\end{equation}

\subsubsection{mRC with clustered full data exchange}

Fig. \ref{Special_case} (d) illustrates a $K$-cluster mRC with each cluster containing $K^\prime$ users. Cluster based mRC allows the users in the same cluster to exchange messages with each other. Full data exchange means that one user in a cluster sends a common message to the remaining $K^\prime-1$ users. Take $K=2$ and $K^{'}=3$ as an example. We denote the messages transmitted in cluster $i$ by $\{s_{i,1}, s_{i,2}, s_{i,3}\}$ where $s_{i,j}$ denotes the message sent from user $j$ in cluster $i$.
The uplink and downlink indicator matrices in cluster $i$ are given by
\begin{equation}\label{System-8}
{\bf D}_{\uparrow} =\bordermatrix{%
& s_{1,1} & s_{1,2} & s_{1,3} & s_{2,1} & s_{2,2} & s_{2,3}  \cr
{\rm user}_{1,1} &                1 & 0 & 0 & 0 & 0 & 0 \cr
{\rm user}_{1,2} &                0 & 1 & 0 & 0 & 0 & 0 \cr
{\rm user}_{1,3} &                0 & 0 & 1 & 0 & 0 & 0 \cr
{\rm user}_{2,1} &                0 & 0 & 0 & 1 & 0 & 0 \cr
{\rm user}_{2,2} &                0 & 0 & 0 & 0 & 1 & 0 \cr
{\rm user}_{2,3} &                0 & 0 & 0 & 0 & 0 & 1 \cr
}
\end{equation}
and
\begin{equation}\label{System-88}
\begin{split}
& {\bf D}_{\downarrow} =\\
& \bordermatrix{%
&  {\rm user}_{1,1} &  {\rm user}_{1,2} &  {\rm user}_{1,3} &  {\rm user}_{2,1} &  {\rm user}_{2,2} &  {\rm user}_{2,3}  \cr
s_{1,1} &                0 & 1 & 1 & 0 & 0 & 0 \cr
s_{1,2} &                1 & 0 & 1 & 0 & 0 & 0 \cr
s_{1,3} &                1 & 1 & 0 & 0 & 0 & 0 \cr
s_{2,1} &                0 & 0 & 0 & 0 & 1 & 1 \cr
s_{2,2} &                0 & 0 & 0 & 1 & 0 & 1 \cr
s_{2,3} &                0 & 0 & 0 & 1 & 1 & 0
}
\end{split}
\end{equation}
where ${\rm user}_{i,j}$ denotes the user $j$ in cluster $i$.
As a result, we have
\begin{equation}\label{System-888}
{\bf D}_{\uparrow} {\bf D}_{\downarrow} =
 \left[
              \begin{array}{cccccc}
                0 & 1 & 1 & 0 & 0 & 0  \\
                1 & 0 & 1 & 0 & 0 & 0\\
                1 & 1 & 0 & 0 & 0 & 0 \\
                0 & 0 & 0 &  0 & 1 & 1 \\
                0 & 0 & 0 & 1 & 0 & 1 \\
                0 & 0 & 0 & 1 & 1 & 0
              \end{array}
            \right].
\end{equation}

\subsubsection{Three-user MIMO Y channel with hybrid data exchange}

Fig. \ref{Special_case} (e) illustrates a three-user MIMO Y channel with hybrid data exchange where each user delivers both private data and common data to other users. Suppose that the messages exchanged in the channel are ordered by $\{s_{1,1}, s_{1,2}, s_{2,1}, s_{2,2}, s_{3,1}, s_{3,2}, s_{1}, s_{2}, s_{3}\}$ where $s_{i,j}$ denotes the $j$th private message sent from user $i$, and  $s_{i}$ denotes the common message sent from user $i$. The uplink and downlink indicator matrices are represented by
\begin{equation}\label{System-9}
\begin{split}
& {\bf D}_{\uparrow} = \\
& \bordermatrix{%
&s_{1,1} & s_{1,2} & s_{2,1} & s_{2,2} & s_{3,1} & s_{3,2} & s_{1} & s_{2} & s_{3} \cr
{\rm user}_1 &                1 & 1 & 0 & 0 & 0 & 0 & 1 & 0 & 0  \cr
{\rm user}_2 &                0 & 0 & 1 & 1 & 0 & 0 & 0 & 1 & 0  \cr
{\rm user}_3 &               0 & 0 & 0 & 0 & 1 & 1 & 0 & 0 & 1
}
\end{split}
\end{equation}
and
\begin{equation}\label{System-99}
{\bf D}_{\downarrow} = \bordermatrix{%
& {\rm user}_1 & {\rm user}_2 & {\rm user}_3 & \cr
s_{1,1} &                0 & 1 & 0 \cr
 s_{1,2} &                0 & 0 & 1 \cr
  s_{2,1} &               1 & 0 & 0 \cr
 s_{2,2} &               0 & 0 & 1 \cr
 s_{3,1} &               1 & 0 & 0 \cr
 s_{3,2} &               0 & 1 & 0 \cr
 s_{1} &               0 & 1 & 1 \cr
 s_{2} &               1 & 0 & 1 \cr
  s_{3} &              1 & 1 & 0
}.
\end{equation}
It is interesting to note that
\begin{equation}\label{System-999}
{\bf D}_{\uparrow} {\bf D}_{\downarrow}  = \left[
              \begin{array}{ccc}
                0 & 2 & 2  \\
                2 & 0 & 2  \\
                2 & 2 & 0
              \end{array}
            \right],
\end{equation}
where each $``2"$ in the $(i,j)$th position means that user $i$ sends user $j$ two messages, including one common message and one private message.

\subsection{Relay Strategies}

There are various relay strategies for the relay channel.  We start with four frequently used relay strategies:

\begin{itemize}
    \item Amplify-and-Forward (AF): The relay performs as an amplifier and forwards a linearly transformed version of its received signal. Appropriate scaling on the transmit signal is applied to satisfy the power constraint at the relay. AF relay is simple and easy to implement in practice. However, the amplification operation also increases the noise power at the relay. As a result, AF relaying usually suffers from poor performance at the low signal-to-noise ratio (SNR) region.
    \item Compress-and-Forward (CF): The signal observed at the relay is quantized, re-encoded, and then transmitted to the destination. CF can suppress the uplink noise to some extent, but cannot completely eliminate the noise propagation effect.
    \item Decode-and-Forward (DF): The messages transmitted from the users are decoded at the relay. Then the recovered bits are re-encoded and transmitted from the relay. This operation avoids noise amplification. However, the decoding operation at the relay is usually costly and time-consuming. More importantly, DF suffers from the multiplexing loss problem since a DF relay is required to decode the individual message of each user.
    \item Detect-and-Forward (DeF): To reduce complexity, DeF is used as a variant of DF. In DeF, the relay detects the transmitted signals by exploiting the modulation information and transmit a scaled version of the detected signals. Compared with DF, the difference of DeF is that the channel decoding is not involved.
        %
\end{itemize}

Network coding techniques can be employed to enhance the relay forwarding performance in the mRC.
In a conventional two-way relay channel, four time slots are required to complete a round of information exchange. Network coding reduces the required time slots from four to three \cite{PLNC_1_ZhangACM2006}, where
the relay receives and decodes information from the two users in two time slots, and then the relay conducts network coding and transmits the network-coded message to the two users. At the user ends, self-interference cancelation technique is used to cancel the self messages in network-coded messages to extract the desired information.
PLNC further reduces the required time slots from three to two \cite{PLNC_1_ZhangACM2006}. Specifically, PLNC allows the two users to transmit signals simultaneously and exploits the benefit of network coding naturally inherent in the superimposed EM waves.

The idea of PLNC can be readily generalized to the mRC. In the mRC, differently from the TWRC, the relay may receive signals sent from more than two users. Conventional relay strategies such as AF and DF can be used in the mRC to produce network coded signals at the relay. However, as aforementioned, AF suffers from the noise amplification problem, and DF suffers from the multiplexing loss problem.

As an alternative, compute-and-forward (CoF) relay strategy can be used to realize PLNC more efficiently in the mRC. Without decoding individual messages, the CoF strategy allows the relay to decode only the modulo sum of the messages \cite{Rate_SISO_TWRC_CoF_theor_3_Nazer_2011}. Relay decodes the noisy linear equations constructed by the channel and the transmitted messages. Destinations obtain their desired messages after receiving and solving sufficiently many linear equations.
In this way, the inherent noise amplification of AF and the inherent multiplexing loss of DF can be avoided, leading to significant performance enhancement.
Unlike other schemes that achieves performance limits by using random coding, CoF is established based on some linearly structured codes, e.g., nested lattice codes. Codebook's linear property guarantees that the integer combination of these codewords is still valid codewords.
In short, the CoF relay strategy not only provides anti-noise protection, but also provides an opportunity to exploit interference to gain cooperation benefits.

\section{Information-Theoretic Studies in Multi-way Relay Channel}\label{information_theoretic}
In this section, we survey the studies of the mRC from an information-theoretic perspective. We categorize these studies based on the performance metrics employed in the analysis. We focus on the metrics of channel capacity and degrees of freedom (DoF). Other metrics including bit-error rate, outage probability, and diversity-and-multiplexing tradeoff are also included for discussion.

\subsection{Capacity Analysis}

From information theory, a rate tuple of the mRC is said to be achievable if we can design a transceiver scheme with this rate tuple to ensure that the decoding error probability approaches zero as $T_{\uparrow}$ and $T_{\downarrow}$ tend to infinity. Collecting all achievable rate tuples gives the capacity region of the mRC.

As discussed above, the capacity region of the mRC is generally unknown, even for the simplest two-way relay model. In general, the capacity analysis is carried out from two aspects, i.e., inner bounds and outer bounds. An inner bound is usually the best known achievable rate tuples, while an outer bound is obtained by introducing certain relaxations into the mRC.
%
%
Inner and outer bounds respectively reduce to single-letter lower and upper bounds when all the exchanged messages have an equal rate.
When an inner bound meets an outer bound, it produces the capacity region of the corresponding mRC. Thus, the aim of the capacity analysis is to tighten the inner and outer bounds as much as possible.

A frequently used capacity outer bound can be found with the cut-set theorem \cite{gamal2011net_infor_ther}. But the cut-set outer bound is generally not tight for an mRC. A tighter outer bound can be obtained by using the genie-aided method \cite{wang2012genieChains,wang2014subspaceAlignment}, or by combining the cut-set theorem and the genie-aided method.

\begin{itemize}
\item \textbf{Cut-set bound}: The cut-set bound refers to a bounding technique which uses cuts to relax the channel model and produce an outer bound. By putting certain cuts on the transmission links of the mRC network, the nodes separated by the same cut form a group and can share transmit/receive messages. With this set up, the nodes in a group can be combined and treated as a new node. By using this relaxation, we can convert the mRC into some simple channel models with known capacities. These capacities can be considered as outer bounds of the considered mRC. In general, the cut-set bound perform loosely for the mRC even when the network size is relatively small.
\item \textbf{Genie-aided bound}: Using extra genie messages is another efficient way to produce outer bounds.
Consider that the received signal at the relay is the degraded version of the signals received at the user ends, the basic idea of genie-aided outer bound characterization is to provide some extra genie messages to the relay such that the relay can decode the remaining messages. This relaxation can be translated to a series of information inequalities which can be used to produce an outer bound. To ensure the tightness of the outer bound, the provided genie messages should be kept as less as possible.
\end{itemize}

Given an outer bound of the mRC, we need to explore transmission strategies to approach this bound by designing encoding and decoding strategies, choosing relay strategies and optimizing the power and spectrum resource of the system.

To date,  the mRC is mostly focused on the models, including TWRC, multi-pair TWRC, Y-channel
and their MIMO versions. We summarize the capacity analysis of these models as follows.

\subsubsection{Single-input single-output (SISO) TWRC}

In the SISO TWRC, a single antenna is equipped at all nodes.
Various relaying strategies have been considered in the pursue of the performance limit of the SISO TWRC \cite{Rate_SISO_26_RankovISIT2006, Rate_SISO_TWRC_theoretical_1_Rankov2005, Rate_SISO_27_JSAC2007,Rate_SISO_6_OechteringTIT2008, Rate_SISO_2_SchnurrITW2008, Rate_SISO_TWRC_theoretical_6_Zhang2009,Rate_SISO_TWRC_theoretical_4_Popovski2007,Rate_SISO_2_SchnurrITW2008,Rate_SISO_19_GunduzAllerton2008,Rate_SISO_TWRC_CoF_theor_2_Nazer2007,Rate_SISO_TWRC_CoF_theoretical_1_Wilson_2010,
Rate_SISO_7_NamIIZSC2008,Rate_SISO_9_NamTIT2010,Rate_SISO_TWRC_CoF_theor_3_Nazer_2011,Rate_SISO_TWRC_CoF_practical_4_Hern2011}. To clearly demonstrate the capacity analysis studies of SISO TWRC, we next review the exiting works according to the type of relay strategies.

Traditional AF and DF relay strategies have been used in the SISO TWRC.
In \cite{Rate_SISO_26_RankovISIT2006, Rate_SISO_TWRC_theoretical_1_Rankov2005, Rate_SISO_27_JSAC2007}, the authors compared the achievable sum rate of the SISO TWRC when using traditional AF and DF relay strategies. Here the DF relay strategy requires the relay to decode both messages from the two source nodes in the first phase. After the two decoded messages being separately re-encoded, a superposition of them is broadcast in the second phase. The results in \cite{Rate_SISO_26_RankovISIT2006, Rate_SISO_TWRC_theoretical_1_Rankov2005, Rate_SISO_27_JSAC2007} show that for general case, the DF relay strategy performs better than the AF strategy due the noise amplification at the AF relay. However, for certain special cases where the relay locates in the middle of the two source nodes, an opposite result can be observed. The authors in  \cite{Rate_SISO_6_OechteringTIT2008,Rate_SISO_2_SchnurrITW2008}
assumed that the relay can successfully decode the two messages of the sources nodes using the DF relay strategy, and then a re-encoded composition message is broadcast by the relay node. They developed the optimal coding strategy with which the capacity region of the BC phase is characterized. Their results show that the XOR superposition approach is in general inferior and only achieve the capacity in some special scenarios. \cite{Rate_SISO_TWRC_theoretical_6_Zhang2009} showed that the cut-set bound of the TWRC can be approached by using DF relaying in the low SNR regime. The used DF relaying scheme assumed to use two ``separated" multiple access to individually decode two messages in the uplink phase, while in the downlink phase, two decoded messages are combined by using physical-layer network coding.

Network-coded DF relaying is an alternative DF based relay strategy. Different from traditional DF relaying, network-coded DF does not demand the decoding of two source messages at the relay but the received signal at the destinations requires \cite{Rate_SISO_TWRC_theoretical_6_Zhang2009,Rate_SISO_19_GunduzAllerton2008}. The network-coded DF relaying decodes the network-coded symbol. Hence, in many research papers, network-coded DF relaying is also called partial DF relaying.  Authors in  \cite{Rate_SISO_19_GunduzAllerton2008} derived an upper bound of the achievable rate for the network-coded DF.
Authors in \cite{Rate_SISO_TWRC_theoretical_6_Zhang2009} proved that the capacity can be achieved at high SNR using the network-coded DF relaying strategy.

The CF strategy requires the relay to quantize its received signals, which are then transmitted to the destination. CF relaying suffers from the quantization noise. Compared with AF relaying, although CF relaying also suffers from noise propagation, it avoids noise amplification.
The authors in \cite{Rate_SISO_2_SchnurrITW2008} proposed a new relay scheme to combine DF relaying and CF relaying. The relay partially decodes the messages and then forwards the decoded information to the destinations using the coding scheme in \cite{Rate_SISO_6_OechteringTIT2008}; the missing information is transmitted to the destinations using a CF strategy. An achievable rate region for the two-phase TWRC with a half-duplex relay node was derived.
The authors in \cite{Rate_SISO_19_GunduzAllerton2008} derived an achievable rate region based on a combination of network-coded DF relaying and CF relaying in which the relay partially decodes the messages and forwards the quantized residue of its received signal to the users together with the decoded part.
The proposed hybrid scheme in \cite{Rate_SISO_19_GunduzAllerton2008} achieves a larger achievable rate region.

The CoF relay strategy allows the relay to produce an appropriate physical-layer network-coded message functions,
which contains just-enough information for each user to recover the other user's message. Thus, the noise amplification of the AF and the multiplexing loss of the DF can be avoided, leading to a significant performance enhancement. CoF relies on structured codes such as nested lattice codes.
In \cite{Rate_SISO_TWRC_CoF_theoretical_1_Wilson_2010}, the authors proposed to utilize the nested lattice codes for the Gaussian TWRC. A rate of ${1\over 2} log(1+{\rm snr})$ bits per transmitter is achieved where ${\rm snr}$ denotes the SNR of the inputs.
%
%
However, the results in \cite{Rate_SISO_TWRC_CoF_theoretical_1_Wilson_2010} were obtained assuming with a symmetric channel, that is, the transmit powers and the noise variances are the same at all nodes. The symmetric channel allows to use the same nested lattice code at the two sources.
%
In \cite{Rate_SISO_7_NamIIZSC2008} and \cite{Rate_SISO_9_NamTIT2010}, the authors considered different transmit power constraints of TWRC.
Different from traditional nested lattice codes, the authors proposed to use a lattice partition chain to generate two different shaping lattices at the two sources. It was shown that the obtained achievable rate region is able to achieve the capacity region within 1/2 bit.
%
%
To avoid using impractical high dimensional lattice codes,
the authors in  \cite{Rate_SISO_TWRC_CoF_practical_4_Hern2011} proposed a CoF scheme based on multilevel coding. Their proposed scheme uses practical linear codes and permits the relay to recover a class of functions. Here the functions are determined by the channel realizations. This flexibility ensures that the proposed scheme outperforms the one using fixed functions.

\subsubsection{Other SISO channels}
The capacity analysis has been extended from TWRC to various other channel models, such as the multi-pair TWRC, the Y channel and the mRC with full data exchange. The data exchange flows of the Y-channel and the mRC with full data exchange are described by eqns. \eqref{System-4}-\eqref{System-44} and eqns. \eqref{System-8}-\eqref{System-88}, respectively.
\begin{itemize}
    \item \textbf{Multi-pair TWRC:}
%
    In \cite{Rate_SISO_10_Avestimehr_2009}, the capacity region was investigated for the deterministic
multi-pair TWRC, which eliminates the channel noise and focuses on the interaction between signals. They showed that the cut-set upper bound can be reached for both full-duplex relay model and half-duplex relay model by using the equation-forwarding relay strategy.
The authors in \cite{Rate_SISO_4_SezginISIT2009} studied the achievable rate region of the Gaussian two-pair full-duplex TWRC. A hybrid transmission scheme by combining both the nested lattice codes and the random Gaussian codes was proposed to achieve the capacity region within 2 bits/sec/Hz per user for arbitrary channel gains.
%
%
%
In \cite{Rate_SISO_14_SezginTIT2012}, the capacity region characterization of the deterministic multi-pair TWRC was extended to a scenario with an arbitrary number of user pairs.
A divide-and-conquer (DC) relaying strategy was proposed to achieve the cut-set bound. The key idea of the DC relaying strategy used to achieve the capacity region is to divide the relay's signal space into multiple orthogonal subspaces and the each user pair transmits on an assigned subspace. It was shown that for the two user-pairs with arbitrary channel gains, the DC relaying strategy achieves the capacity region within 3 bits/sec/Hz per user.
%
%
%
Then, the authors in \cite{Rate_SISO_MultiPairTWRC_NEW_1_Li2017, Yuan_TIT_2019} studied the achievable rate of the Gaussian two-pair TWRC with any channel conditions. It was showed that the capacity can be reached within 1/2 bit by using the proposed message-reassembling strategy.
%

    \item \textbf{Y channel:}
A. Chaaban \emph{et al.} in \cite{Rate_SISO_23_ChaabanISIT2011, Rate_SISO_18_ChaabanAsilomar2011, Rate_SISO_5_ChaabanTIT2013, Rate_SISO_20_ChaabanTIT2015,Rate_SISO_Y_Channel_new_1_Ibrahim2018} analyzed the capacity of the Y channel. They first showed that the cut-set bounds are insufficient to characterize the capacity region of the Y channel.
By developing new tighter upper bounds, the capacity region was determined for the deterministic Y-channel in \cite{Rate_SISO_23_ChaabanISIT2011, Rate_SISO_20_ChaabanTIT2015}.
In \cite{Rate_SISO_18_ChaabanAsilomar2011},
the authors analyzed the achievable sum-rate using the functional DF relay strategy for the Gaussian 3-user Y channel.
The constant gaps of the upper and lower additive/multiplicative bounds were characterized.
The authors showed that when all nodes have equal powers, the corresponding additive gap and multiplicative gap are less than 2 bits and 4 bits, respectively.
%
%
The authors in \cite{Rate_SISO_5_ChaabanTIT2013} extended the result in \cite{Rate_SISO_18_ChaabanAsilomar2011} to a $K$-user case, and proved that the sum-capacity can be achieved within $2\log(K-1)$ bits.
In \cite{Rate_SISO_20_ChaabanTIT2015}, the authors extended the result obtained for the deterministic Y channel to the Gaussian Y channel by including superposition coding, nested-lattices, and successive decoding as key components. The proposed scheme was shown to be capable of achieving the capacity region within a constant gap independent of the channel conditions.
%
%
The authors in \cite{Rate_SISO_Y_Channel_new_1_Ibrahim2018} analyzed the achievable rate of the deterministic Y-channel with a hybrid data exchange model, that is, a user transmits private and common messages simultaneously to other two users. Genie aided technique and cut-set theorem were used to produce a tighter outer bound. The authors showed that under certain special scenarios, the obtained outer bound can be achieved.

\item  \textbf{mRC with full data exchange:} 
The authors in \cite{Rate_SISO_25_GunduzTIT2013} studied a generalized multiway relay channel. They characterized the achievable rate regions for the Gaussian mRC by considering a full data exchange model. The authors demonstrated that the capacity region can be achieved with a constant gap by using the CF relay strategy independent of the power setting of all nodes. Moreover, the AF relay strategies were proven to achieve the capacity within a finite-bit gap.
%
\end{itemize}

\subsubsection{MIMO TWRC}

With multiple antennas, a MIMO TWRC can be formed to further improve the system performance via spatial resource utilization. Exiting studies on the capacity analysis can be roughly classified into two directions, where the first one focused on the capacity analysis of the broadcast phase by assuming that the messages have been successfully decoded in the first phase; the second direction considered effect of transmissions over two phases and attempted to design efficient PLNC operations at the relay to achieve a higher rate. We next review the existing studies in these two directions.

Under the assumption that the relay decodes the messages sent from two sources, the uplink channel of the MIMO TWRC reduces to a classical MIMO MAC channel.
Then, the research effort was focused on the downlink phase.
The authors in \cite{Rate_MIMO_4_WyrembelskiISIT2008} determined the capacity region of the BC phase in the MIMO TWRC. The authors in \cite{Rate_MIMO_9_OechteringTCOM2009} analyzed the capacity region of the BC phase in the discrete memoryless MIMO Gaussian TWRC. The capacity region problem was characterized by establishing a weighted sum rate optimization problem, which is solved by using the iterative fixed point algorithm.
%
%
The authors in \cite{Rate_MIMO_6_WyrembelskiTWC2011} extended the results in \cite{Rate_MIMO_4_WyrembelskiISIT2008} and studied the capacity region of the BC phase in the MIMO Gaussian TWRC with additional common messages. The authors transformed the capacity characterization problem to a transmit covariance matrix optimization problem.
%

Decoding both messages at the relay, however, suffers from multiplexing loss as compared with PLNC techniques. We next describe several relaying schemes to support PLNC in MIMO TWRC.
The basic idea is to apply the generalized singular-value-decomposition (GSVD) technique to jointly parallelize the channel matrices associated with different users. PLNC is then used at each parallel scalar channel.
Assume that $n_A$, $n_B$, and $n_R$ represent the antenna numbers at user $A$, user $B$, and the relay, respectively. For $\max(n_A, n_B) \geq n_R$, it was showed that the capacity of the MIMO TWRC is asymptotically reached at high SNR \cite{Rate_MIMO_3_YangTIT2011}.
In \cite{Rate_MIMO_1_YangTCOM2013}, the authors further developed an eigen-direction alignment-based scheme to improve performance at medium and low SNR scenarios.

For configurations with $\max(n_A, n_B) < n_R$, the schemes in \cite{Rate_MIMO_3_YangTIT2011} and \cite{Rate_MIMO_1_YangTCOM2013} in general perform worse and the obtained achievable rate is far away from the capacity.
To address this challenge, a cooperative precoding scheme with reduced dimension was proposed in \cite{Rate_MIMO_10_YangTWC2012} for the MIMO TWRC. The authors proved that the sum-capacity is asymptotically achieved within $1/2$ bit per transmit antenna.
The authors in \cite{Rate_MIMO_7_YuanTIT2013} developed a scheme combing network-coding and space-division for the MIMO TWRC.
The authors demonstrated that the sum-capacity is asymptotically achieved at high SNR within 0.161 bits per user-antenna, for arbitrary antenna configurations and arbitrary channel conditions. The large-system analysis over Rayleigh-fading channels showed that the average asymptotic sum-capacity is achieved within less than 0.053 bits per relay-antenna.

\subsubsection{Other MIMO models}
The capacity analysis has been extended to other channel models, including the MIMO multi-pair TWRC, the MIMO cellular TWRC, and the mRC with full data exchange. The data exchange flows of the cellular TWRC are described by eqns. \eqref{System-6} and \eqref{System-66}.

\begin{itemize}
    \item \textbf{MIMO multi-pair TWRC:}
    As an extension of \cite{Rate_MIMO_7_YuanTIT2013}, the authors in \cite{Rate_MIMO_8_XinTSP2016} analyzed the MIMO multi-pair TWRC from a new principal-angle perspective. The basic idea is to use principal angles as a measure to determine a dimension ratio of two partitioned subspaces, the PLNC decoding subspace and the complete decoding subspace, from the relay's signal space.
    %
    Then the user precoders and the relay processing matrix were optimized with an aim to maximize the asymptotic sum-rate at high SNR. The authors demonstrated that the proposed scheme works better than the existing counterpart schemes and has the minimum gap with the cut-set bound.

    \item \textbf{MIMO cellular TWRC: }
    The MIMO cellular TWRC was studied in \cite{Rate_MIMO_Cellular_1_Ding2011,Rate_MIMO_Cellular_2_Sun2012,Rate_MIMO_Cellular_3_Chiu2012} for AF relaying strategy.
    In \cite{Rate_MIMO_Cellular_1_Ding2011}, linear precoding at the BS was designed to realize signal alignment. In specific, the BS precoder aims to align each signal stream transmitted from the BS to the direction of the signal transmitted from a user which the BS intends to exchange with.
    In \cite{Rate_MIMO_Cellular_2_Sun2012} and \cite{Rate_MIMO_Cellular_3_Chiu2012}, linear precoder at the BS and the processing matrix at the relay were optimized according to various criteria, including sum-rate maximization and max-min signal-to-interference-plus-noise ratio (SINR).
    It is noted that AF relaying inherently suffers from relay noise propagation and inefficient relay power utilization. To address this issue, DF-based relaying scheme is used in \cite{Rate_MIMO_2_YangJSAC2012} for the MIMO cellular TWRC. More specifically, linear precoding, dirty-paper coding and nested lattice coding were respectively used at the BS, the relay and the users, respectively. The authors showed that the proposed scheme significantly outperforms AF relaying in \cite{Rate_MIMO_Cellular_1_Ding2011,Rate_MIMO_Cellular_2_Sun2012,Rate_MIMO_Cellular_3_Chiu2012}, but it still cannot reach the capacity especially when with a relatively large MIMO size.
The authors in \cite{Rate_MIMO_11_FangTSP2014} proposed to use a nonlinear lattice precoding scheme in the uplink phase, and produced a nested-lattice-coding assisted PLNC scheme to maximize the sum-rate of the MIMO cellular TWRC. By combining linear precoding and successive interference cancellation, the authors characterized the achievable rate and established the sufficient conditions under which the sum capacity of the system can be asymptotically achieved at high SNR.

\item \textbf{mRC with full data exchange: }

The authors in \cite{Rate_MIMO_New_3_Yuan2018} investigated the MIMO mRC with distributed relays under the full data exchange mode.
The authors proposed a PLNC-based scheme by jointly considering the linear precoding, the nested lattice coding and the lattice-based precoding, which were applied to realize the functions of signal alignment, PLNC and interference mitigation, respectively.
%
The authors showed that the sum capacity can be asymptotically achieved within a constant gap at high SNR.
The authors in \cite{Rate_MIMO_New_4_Lu2018} investigated the sum-rate analysis of the mRC with AF relaying and full data exchange strategy.
The authors analyzed the sum-rate lower and upper bounds. Specifically, a closed-form low bound of the average sum-rate was obtained when the user number is sufficient large. 
In \cite{Rate_MIMO_New_2_Li2018}, the authors considered an mRC with a
multi-antenna relay. The relay beamforming matrices and users' linear receiving
processing were jointly designed with an aim to maximize the
minimum SINR subject to the relay power constraint.
\end{itemize}

\subsection{DoF Analysis}

As discussed above, the capacity of the mRC is generally unknown, even for the simplest two-way relay model. Most existing work was focused on establishing tighter achievability and converse bounds.
These bounds generally becomes lose for relatively complicated mRC models, especially when the user number and the antenna number at each node become large.
%
Due to the difficulty of the rate gap analysis, the DoF has been considered as an alternative metric which has been widely used to characterize the performance limit of the mRC. Roughly speaking, the DoF is the slope
of the capacity against the SNR of the channel as SNR goes to infinity.
The DoF of a channel can also be interpreted as the number of independent data streams which are reliably transmitted over the channel.
The DoF has been widely used to analyze the performance the MIMO mRC.

%


We start with a brief discussion of the techniques frequently used in the DoF analysis. In general, the DoF analysis is performed from two aspects, i.e., achievability and converse. The achievability says that a certain DoF is achievable, while the DoF converse gives certain DoF upper bounds. When the achievablity meets the converse, we claim that the DoF of the corresponding mRC is obtained.


For the converse analysis, two commonly used approaches are the \emph{cut-set bound} and the \emph{genie-aided bound}. The basic ideas of two bounds are similar to the ones introduced in Subsection A, and thus are omitted here for brevity. We next focus on the DoF achievability techniques, as detailed below.



\begin{itemize}
  \item \textbf{Signal alignment}\cite{DoF_Linear_27_LeeTIT2010, DoF_Linear_3_LiuICC2015, DoF_Linear_11_LeeTWC2012,DoF_Linear_31_LiuTIT2015, DoF_Linear_34_WangTSP2014}: Signal alignment allows the data streams exchanged between a certain subsets of users are aligned to occupy a low-dimensional receiving subspace at the relay. Signal alignment is achieved by designing transmit beamformers at the users. Although the data streams aligned in the aligned subspace are not decodable at the relay, they can be decoded at the user ends by using self-interference cancelation. Take signal alignment between two users as an example. The signals transmitted from the two users can be aligned into a common direction such that they only occupy a one-dimension subspace at the relay's receiving space. Then, signal nulling can be applied at the relay for decoding a network-coded signal. Thus, the main function of the signal alignment is to more efficiently use the relay's signal space to include more data streams, so that a higher DoF can be achieved.


  \item \textbf{Interference alignment}\cite{DoF_Linear_15_WangISIT2014, DoF_Linear_2_WangeUC2013}: The goal of interference alignment is to align interference signals at a desired user end such that they occupy a low-dimensional subspace while leaving more dimensions for delivering desired data streams. Interference alignment can be used to enhance the achievable DoF in the clustered based mRC. Since the signals exchanged in one cluster are interference to the users in other clusters, we can use interference alignment to suppress the interference dimension in the downlink transmission to improve the performance.



  \item \textbf{Uplink downlink symmetry}: Symmetric beamforming design for the upplink and the downlink is an important technique in proving the DoF achievability for the mRCs with pairwise data exchange \cite{ DoF_Linear_34_WangTSP2014, DoF_Linear_32_WangTIT2016, DoF_Linear_26_WangJSAC2015}. The upplink/downlink symmetry means that if the signal design, including user beamforming design and relay receiving matrix design, at the uplink can ensure the signal decodability at the relay node, a symmetric beamforming design for the downlink always exists to guarantee the signal decodability at the user ends. We note that the upplink/downlink symmetry exists for the pairwise data exchange but not for other data exchange patterns.

\item \textbf{Solving nonlinear equations}: The joint precoding design at the users and the relay is usually a highly nonlinear problem. A frequently used approach is to decouple the overall design problem into two separate subproblems, where one subproblem is for the uplink and the other one is for the downlink. Then, signal signal and interference alignment techniques can be used to meet certain linear constraints in the precoding design \cite{DoF_Linear_32_WangTIT2016}. However, the optimality of this decoupling approach is generally unknowns, except for some simple mRC topologies with uplink downlink symmetry. As such, a more general approach to the DoF analysis of the MIMO mRC involves solving a set of nonlinear equations. Advanced matrix manipulation techniques are usually required in finding good solutions to the system of nonlinear equations.

\end{itemize}

We now discuss the DoF results for various MIMO mRC models, including the MIMO Y channel, the MIMO multipair TWRC, the MIMO two-way X relay channel, and the cluster-based mRC. The data exchange flows of the the two-way X relay channel are described by eqns. \eqref{System-7} and \eqref{System-77}.

\subsubsection{DoF of MIMO Y Channel}

In a $K$-user MIMO Y channel, each user intends to exchange private messages with the remaining $K-1$ users as specified in Section II.

The authors in \cite{DoF_Linear_4_LeeISIT2009, DoF_Linear_27_LeeTIT2010} investigated the DoF of the three-user MIMO Y channel assuming that the antenna setting is symmetric, that is, $M$ antennas are equipped at all users and $N$ antennas are equipped at the relay.
For the two users who exchange messages,
the signal alignment technique was used to suppress their uplink signals into a common subspace. The suppressed signals of these two users can be regarded as a network-coded signal. Then in the downlink, the authors proposed to use the network-coding aware interference nulling technique to ensure that all user ends have enough space to decode the desired signals. By utilizing the above two techniques, the authors demonstrated that the optimal DoF of $3M$ is achieved when $N\geq \lceil\frac{3M}{2}\rceil$.

In \cite{DoF_Linear_50_DingTWC2017}, the authors also investigated the DoF of the three-user MIMO Y channel. Different from \cite{DoF_Linear_4_LeeISIT2009, DoF_Linear_27_LeeTIT2010}, the authors considered using $K$ geographically separated relay nodes to help signal forwarding. The signal alignment and interference nulling techniques are not applicable here due to the fact that joint signal processing of the geographically separated relays is not possible. Instead, the authors proposed a novel framework to characterize the achievable DoF.
The framework transforms the DoF achievability to the solvability of a nonlinear equation problem with user precoding matrices, user receiving matrices, and relay precoding matrices as unknown variables.
To address the solvability of the problem, the authors proposed a general method for finding the solution of the nonlinear equations. The obtained results showed that for $K$ $N$-antenna relay nodes, the optimal DoF is achieved for $\frac{M}{N}\in [0, \max\{\sqrt{\frac{3K}{3}, 1} \}\cup [\frac{3K+(9K^2-12K)^{1/2}}{6} \infty)$.

In \cite{DoF_Linear_42_Chaaban2013}, the authors investigated the DoF of three-user MIMO Y channel. Different from \cite{DoF_Linear_4_LeeISIT2009, DoF_Linear_27_LeeTIT2010}, the authors in \cite{DoF_Linear_42_Chaaban2013} considered a more general case where the users are equipped with arbitrary number of antennas. Let $M_1$, $M_2$ and $M_3$ be the antenna numbers equipped at the three users and $N$ antennas at the relay.  It was shown that $\min\{2M_2+2M_3, M_1+M_2+M_3,2N\}$ DoF can be achieved.

In \cite{DoF_Linear_45_MaierarXiv2014}, the authors studied the DoF of the three-user single-antenna Y channel. To derive the optimal DoF, the authors developed optimal cyclic interference alignment schemes for the channel with conceptual polynomial channel model. 

The authors in \cite{DoF_Linear_10_YuanCL2014} extended the three-user MIMO Y channel considered in \cite{DoF_Linear_4_LeeISIT2009, DoF_Linear_27_LeeTIT2010} to the four-user MIMO Y channel. In addition to use the signal alignment technique in \cite{DoF_Linear_4_LeeISIT2009, DoF_Linear_27_LeeTIT2010}, a new technique termed unpaired signal alignment was developed to more efficiently use the relay's signal space especially when the intersection space of a user-pair does not exist. The per-user DoF of $\max \{\min(M, \frac{3}{8}N), \min(\frac{6}{7}M, \frac{N}{2})\}$ is proven to be achievable by using the proposed scheme. Unfortunately, the optimality of the derived DoF still remains an open problem.

The $K$-user symmetric MIMO Y-channel was studied in \cite{DoF_Linear_48_ChaabanTIT2015}. The authors proposed a two-step strategy, i.e., channel diagonalization and cyclic communication, to achieve the optimal DoF. Zero-forcing precoding was employed to diagonalize and parallelize the channel into multiple sub-channels. Cyclic communication and signal-space alignment were applied at each sub-channel. It was shown that the optimal DoF region of the channel can be achieved when $N \leq M$.

By considering the asymmetric $K$-user channel model where $M_i$ antennas are equipped at user $i$, the authors in \cite{DoF_Linear_28_LeeICC2011, DoF_Linear_11_LeeTWC2012} claimed that a total DoF of $dK(K-1)$ is achieved if $M_i\geq d(K-1)$, $N\geq \frac{dK(K-1)}{2}$ and $N<\min\{M_i+M_j-d \mid \forall i\neq j\}$ where $d$ denotes the average per-user DoF.
In establishing the results, the signal alignment technique was used in both the uplink and the downlink.

In \cite{DoF_Linear_12_LeeCL2013}, the authors investigated a hybrid transmission in MIMO Y channel, i.e., both private messages and common messages are transmitted over the network. The authors derived the antennas requirements at the relay when each user obtains one DoF for each type message.

In \cite{DoF_Linear_37_LeeICC2011, DoF_Linear_17_LeeTIT2014}, the authors derived an achievable DoF of the $K$-user MIMO Y channel with multicast transmission, that is, each user sends common messages to all remaining $K-1$ users. By assuming that $M$ antennas are equipped at all nodes, the authors showed that the optimal total DoF of $\frac{KM}{K-1}$ is achieved when all nodes have knowledge of global CSI and operate in the
full-duplex mode. The achievability was established by using a repetition coding scheme.
Furthermore, under the assumption that all nodes only have local CSI,
a total DoF of $\frac{K}{2}$ is shown to be achievable when $M=K-1$. PLNC and successive interference cancelation were used to achieve this DoF.

In \cite{DoF_Linear_24_LeeISIT2013,DoF_Linear_35_LeeArXiv,DoF_Linear_39_LeeGC2014,DoF_Linear_40_LeeJSAC2015}, the authors studied the MIMO Y channel with fully-connected communications, where the direct links between users are also considered in the design. The authors showed that a sum-DoF of $\frac{K(K-1)}{2K-2}$ is optimal if the relay has $N$ antennas with $N\geq K-1$. The authors also proved that the DoF of the fully-connected network cannot be improved even if the relay has an infinite number of antennas.

In \cite{DoF_Linear_25_HuangGC2014}, the authors studied the DoF of the half-duplex MIMO Y channel with a common data exchange. The authors found that different from the private data exchange, the optimal DoF can be reached with asymmetric traffic load in the MAC phase and the BC phase. By optimizing the time allocation between the MAC phase and the BC phase, it was shown that the total DoF of the channel with asymmetric traffic load is significantly higher than the channel with equal time allocation between the two phases.

In \cite{DoF_Linear_23_SalahAsilomar2015}, the authors studied the DoF of the MIMO Y channel with a hybrid data exchange model by assuming that each user conveys $K-1$ private messages and one common message to the other users. By using the partial signal alignment for the common messages in the uplink phase and the zero-forcing precoding at the downlink phase, the authors showed that the optimal total DoF of $K\min(N,M)$ can be reached.

\begin{figure}[tp]
\begin{centering}
\includegraphics[scale=0.45]{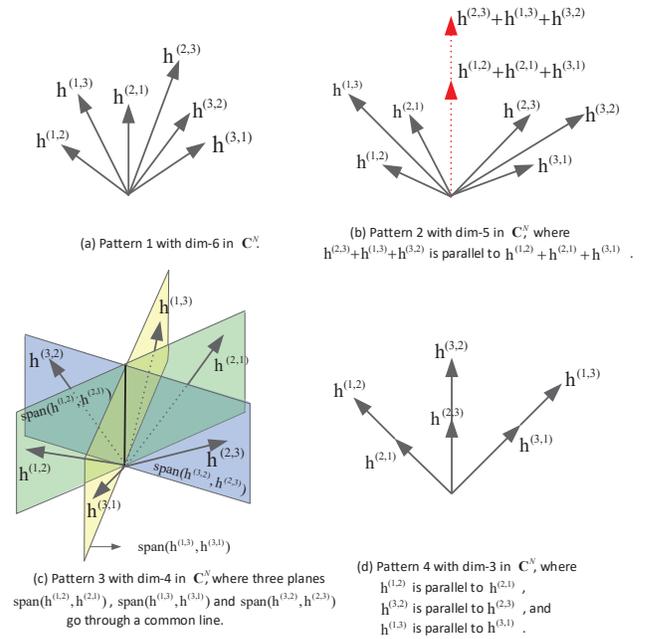}
\vspace{-0.1cm}
\caption{A geometric illustration of Patterns  for three-user MIMO Y channel.}  \label{Fig_Pattern_L_1}
\end{centering}
\vspace{-0.3cm}
\end{figure}

In \cite{DoF_Linear_46_WangGC2014} and \cite{DoF_Linear_34_WangTSP2014}, the authors proposed a systematic approach to analyze the achievable DoF of the MIMO Y channel with an arbitrary number of users. It is assumed that each user has $M$ antennas and the relay has $N$ antennas. To characterize the DoF analysis, the authors first proved the DoF symmetry between the MAC phase and the BC phase. This symmetric property means that if the network-coded signals are decodable at the relay node, the desired signals can also be decoded at receiver ends.
%
%
The main idea of the systematic approach is first to divide all the data streams transmitted over the channel into a number of unit. We require that in each unit, only one data stream from a user is included. The subspace spanned by a unit is called a pattern. The key step of the approach is to design a pattern using different types of signal alignment techniques such that data streams in a unit occupies the smallest dimension subspace in the relay's receiving space. Here we need to guarantee that all data streams in one unit are decodable in the desired receiver with the known self-interference. Meanwhile, the subspace occupied by all units are orthogonal to each user. After that, via counting the unit number with the units being able to be supported by the relay's space,
we can finally determine the total number of data streams which can be delivered by the channel. Take the three-user MIMO Y channel as an example. Assume that ${\bf h}^{k,k^\prime}$ denotes the signal direction of the data streams transmitted from user $k$ to user $k^\prime$ and $\mathcal{U}$ as the subspace spanned by a pattern. Four different patterns can be constructed as shown in Fig. \ref{Fig_Pattern_L_1}. The efficiency and the antenna configuration requirements of these patterns are demonstrated in TABLE \ref{Table OneCluster}, the ``dimension" presented in the second the column implies the dimension number of the subspace occupied by a pattern; $d_{\mathrm{sum}}$ in the third column denotes the total number of data streams we pack in one unit; $d_{\mathrm{relay}}$ in the fourth column define the ratio of the ``dimension" and $d_{\mathrm{sum}}$ which actually determine the efficiency of the pattern; the ``requirement" presented in the fifth column implies the antenna configuration required to construct the pattern.

%

\begin{table}[!t]
\centering
\caption{Patterns for the three-user MIMO Y channel}
\label{Table OneCluster}
\begin{IEEEeqnarraybox}[\IEEEeqnarraystrutmode\IEEEeqnarraystrutsizeadd{2pt}{1pt}]{v/c/v/c/v/c/v/c/v/c/v}
\IEEEeqnarrayrulerow\\
&\mbox{Pattern}&&\mbox{Dimension}&& d_{\mathrm{sum}} && d_{\mathrm{relay}} && \mbox{Requirement} &\\
\IEEEeqnarraydblrulerow\\
\IEEEeqnarrayseprow[3pt]\\
& 1 && 6 && 6 && 1 && \frac{M}{N}>0 &\\
\IEEEeqnarrayseprow[3pt]\\
\IEEEeqnarrayrulerow\\
\IEEEeqnarrayseprow[3pt]\\
& 2 && 5 && 6 && \frac{6}{5} && \frac{M}{N}>\frac{1}{3}&\\
\IEEEeqnarrayseprow[3pt]\\
\IEEEeqnarrayrulerow\\
\IEEEeqnarrayseprow[3pt]\\
& 3 && 4 && 6 && \frac{3}{2} && \frac{M}{N}>\frac{1}{3}&\\
\IEEEeqnarrayseprow[3pt]\\
\IEEEeqnarrayrulerow\\
\IEEEeqnarrayseprow[3pt]\\
& 4 && 3 && 6 && 2 && \frac{M}{N}>\frac{1}{2}&\\
\IEEEeqnarrayseprow[3pt]\\
\IEEEeqnarrayrulerow
\end{IEEEeqnarraybox}
\end{table}

The authors in \cite{DoF_Linear_3_LiuICC2015} further improved the results obtained in \cite{DoF_Linear_46_WangGC2014, DoF_Linear_34_WangTSP2014} by developing a generalized signal alignment scheme, 
where a compression relay matrix was used to first reduce the dimension at the relay node to make the signal alignment feasible. The authors shown in \cite{DoF_Linear_3_LiuICC2015} that the generalized signal alignment achieves a broader DoF region.

In \cite{DoF_Linear_6_MuTWC2014}, the authors proposed a signal-group based alignment strategy to achieve the optimal total DoF for MIMO Y channel. The basic idea of the signal-group alignment is similar to the unit partition based methods developed in \cite{DoF_Linear_46_WangGC2014} and \cite{DoF_Linear_34_WangTSP2014}. That is, when the requirement on the antenna configuration is not met for the two-user signal alignment,
the signal alignment of more users can be performed to reduce the size of the occupied subspace in the MAC phase.

In \cite{DoF_Linear_44_WangICC2015}, the authors investigated the secrecy DoF characterization of the MIMO Y channel with unicast data exchange mode. Utilizing the notion of orbit original defined in abstract algebra, the authors categorized the transmission patterns into a couple of orbits. Here an orbit refers to a minimum subset of users whose data exchanges are limited to this subset. The achievable secrecy DoF of the channel with different numbers of orbits was analyzed. The authors proved that the DoF capacity is achievable when with one and two orbits.

The work in \cite{DoF_Linear_14_ZewailICC2014, DoF_Linear_1_ChaabanEuropean2014,DoF_Linear_36_ChaabanITG2015, DoF_Linear_49_LiuCL2016} investigated the DoF region of the MIMO Y channel. In general, it is more difficult to characterize the DoF-region than the total DoF since the DoF-region characterization involves analyzing the asymmetric data
transmissions, which necessitates the multi-dimensional DoF analysis.

In \cite{DoF_Linear_14_ZewailICC2014}, the authors characterized the DoF region of the three-user and four-user MIMO Y channel with an arbitrary antenna setting. By using signal alignment and the detour scheme originally proposed for the capacity achieving of the deterministic relay networks, the authors showed that the optimal DoF region can be achieved under certain setting of users' and relay's antennas.

In \cite{DoF_Linear_1_ChaabanEuropean2014,DoF_Linear_36_ChaabanITG2015}, the authors studied the DoF region of the MIMO Y channel with arbitrary users. The authors proposed to simultaneously diagonalize all uplink/downlink channels to derive the achievable DoF. The authors showed that the optimal DoF region is achieved when each user has more antennas than the relay.

The authors in \cite{DoF_Linear_49_LiuCL2016} improved the result in \cite{DoF_Linear_14_ZewailICC2014} and showed that the optimal DoF region of the three-user asymmetric MIMO Y channel is achievable by utilizing antenna deactivation, pairwise and cyclic signal alignments.

\subsubsection{DoF of MIMO multipair TWRC}
Compared with the MIMO Y channel, the MIMO multipair TWRC mainly differs in the data exchange pattern. Given this difference, most of the techniques (such as signal alignment) for the MIMO multipair TWRC have their counterparts for the MIMO Y channel discussed above.

In \cite{DoF_Linear_2_WangeUC2013}, the authors characterized the optimal per-user DoF for the two-pair and three-pair TWRCs with respect to $\frac{M}{N}$. As in \cite{DoF_Linear_46_WangGC2014, DoF_Linear_34_WangTSP2014}, by using different types of signal alignments, the authors showed that the per-user DoF is piecewise linear depending on $M$ and $N$ alternately. In addition, the authors proposed to utilize the linear dimension counting to obtain a linear upper bound.

In \cite{DoF_Linear_13_LeeTWC2013}, the authors studied the MIMO TWRC with two relay nodes. Three achievable schemes, namely, time-division multiple access, signal space alignment for network coding, and interference neutralization, were utilized to characterize the achievable DoF. The authors showed that an DoF of $\max\{\min(4N,2M), \min(2N, 2\lfloor\frac{4}{3}M\rfloor), \min(2N-1,4M) \}$ is achievable.

In \cite{DoF_Linear_16_MaierITW2013}, the authors studied the same channel model with \cite{DoF_Linear_13_LeeTWC2013} but assuming that only one-antenna is equipped at each node. By using the cyclic interference neutralization, the authors demonstrated that the DoF of $4$ is asymptotically achieved if the channel satisfies a certain set of symmetry conditions.

In \cite{DoF_Linear_47_WangICC2015, DoF_Linear_32_WangTIT2016}, the authors investigated the MIMO two-pair TWRC with arbitrary $K$ relays. The authors established a general framework for analyzing the DoF.
The proposed framework involves two key techniques, i.e., signal space alignment and interference neutralization, which are further transformed to joint designs of the user transmit precoding matrices, relay process matrices, and user receive matrices. The transmit precoding matrices and the user receive matrices were first determined to achieve signal alignment within the uplink and downlink for more efficient interference neutralization.
With the determined transmit beamformers and user receive beamformers, the original joint transceiver design boils down to a linear problem only depending on the
relay precoding matrices. The solution is then found by transforming the problem to linear matrix equations. The result in \cite{DoF_Linear_32_WangTIT2016} improves the results obtained in \cite{DoF_Linear_13_LeeTWC2013}. For example, when $K=2$, the optimal DoF is reached for $\frac{M}{N} \in (0,1/2)\cup (1,2)$.

In \cite{DoF_Linear_52_LiuTSP2017}, the authors investigated the DoF region of the MIMO two-pair TWRC considering asymmetric data exchange, that is, the numbers of data streams transmitted from the two users in a pair are different.
Via joint designing the relay compression matrix and user
precoding matrices, the optimal DoF region was established therein.

\subsubsection{DoF of MIMO two-way X relay channel}

Regarding the DoF of the MIMO two-way X relay channel, the authors in \cite{DoF_Linear_18_XiangGC2012, DoF_Linear_19_XiangTSP2013} derived the optimal DoF under certain antenna configurations. In specific, by using two-user signal alignment, the authors showed that the total DoF of $2\min(M,N)$ is achieved for $N\leq \lfloor\frac{8M}{5}\rfloor$.

The DoF of the MIMO two-way X relay channel has also been studied in \cite{DoF_Linear_30_LiuGC2015, DoF_Linear_31_LiuTIT2015}. For the case of $N\geq 2M$, the authors used generalized signal alignment \cite{DoF_Linear_3_LiuICC2015} to perform PLNC. The signals which are exchanged in a pair are aligned in a relay's compressed subspace. With the proposed scheme, the authors showed that the DoF in more broader region of $\frac{M}{N}$ can be achieved. The achievability proof was also extended to other data exchange models such as the MIMO Y channel and the MIMO multipair TWRC.

The approach in \cite{DoF_Linear_49_LiuCL2016} was aslo extended to characterize the optimal DoF region of the MIMO two-way X relay channel \cite{DoF_Linear_51_LiuTCOM2018}.

\subsubsection{DoF of cluster based mRC}

The cluster based mRC refers to a family of multiway relay networks where the users are separated into clusters, and only the users in the same cluster are allowed to exchange messages. The data exchange model in a cluster can be either unicast or multicast.

The authors in \cite{DoF_Linear_21_TianISIT2013, DoF_Linear_20_TianICC2013, DoF_Linear_22_TianTIT2014} first studied the DoF of the cluster based mRC with unicast data exchange in each cluster. One main contribution of this series of work lies in deriving a tighter genie aided upper bound.
For a case of the two-cluster two-user channel model, the authors demonstrated that the optimal DoF is reached for an arbitrary antenna setting at the users and the relay.
For the channel model with arbitrary clusters and users, the conditions of achieving the optimal DoF were established.

In \cite{DoF_Linear_26_WangJSAC2015}, the authors studied the same channel as in \cite{DoF_Linear_21_TianISIT2013, DoF_Linear_20_TianICC2013, DoF_Linear_22_TianTIT2014} with a symmetric channel setting, i.e., the number of antennas at a user is equal to each other. The work in \cite{DoF_Linear_26_WangJSAC2015} improved the DoF obtained in \cite{DoF_Linear_21_TianISIT2013, DoF_Linear_20_TianICC2013, DoF_Linear_22_TianTIT2014}. To achieve this goal, the authors again used the unit/pattern based idea similar to our work in \cite{DoF_Linear_46_WangGC2014, DoF_Linear_34_WangTSP2014} and proposed a systematic approach to design the signal alignments. Different from two-user signal alignment, the proposed systematic signal alignment applies to arbitrary antenna configurations. According to that, achievable DoFs were derived
for the general MIMO mRC. It was showed that the proposed scheme achieves the optimal DoF in a boarder region of $\frac{M}{N}$ than \cite{DoF_Linear_21_TianISIT2013, DoF_Linear_20_TianICC2013, DoF_Linear_22_TianTIT2014}. Moreover, the results treated the optimal DoF obtained in \cite{DoF_Linear_42_Chaaban2013} and \cite{DoF_Linear_15_WangISIT2014} for three-user and four-user MIMO Y channel, and in \cite{DoF_Linear_2_WangeUC2013} for MIMO two-way multipair relay Channel as special cases.

Different from \cite{DoF_Linear_21_TianISIT2013, DoF_Linear_20_TianICC2013, DoF_Linear_22_TianTIT2014, DoF_Linear_26_WangJSAC2015}, the work in \cite{DoF_Linear_33_YuanTIT2014} studied the cluster based mRC with multicast data exchange in each cluster. Similarly to \cite{DoF_Linear_26_WangJSAC2015}, the authors divided the signal streams into multiple units and designed different types of signal alignments to construct the patterns with an aim to suppress the dimensions of the occupied subspace of the signal streams in a unit. The obtained results showed that the proposed scheme achieves the optimal DoF when $\frac{M}{N}\leq \frac{1}{LK-1}$ and $\frac{M}{N}\geq \frac{(K-1)L+1}{KL}$.

\subsubsection{Other channel models}

The work in \cite{DoF_Linear_7_WangTVT2014} studied the cellular TWRC with a hybrid data exchange where the downlink transmission from BS to users is multicast transmission while the the uplink transmissions from the users to the BS are unicast transmissions. By using several techniques including PLNC, signal
alignment, block diagonalization, and the newly proposed
successive interference nulling, the authors proved that the cut-set bound can be reached when cell number is smaller than three. The obtained results revealed that the achievable DoF in cellular TWRC can be doubled under
symmetric antenna configurations; however, the performance is degraded
when antenna configuration is asymmetric.

\subsection{Other Performance Metrics}


There are other metrics to evaluate the performance of the mRC networks. In what follows, we overview the performance studies for the mRC based on other metrics, such as the bit-error-rate, outage probability, and diversity and multiplexing tradeoff (DMT).

\subsubsection{Bit-error-rate}

In \cite{BER_2_IslamIET2013}, the authors studied the error performance of the Y channel with binary phase shift keying modulation. In specific, the authors quantified the possible error events with $L$-user DF and AF relaying. An asymptotic bounds on the probability of that a user incorrectly decodes the messages were derived. The authors demonstrated that the higher order error events occur with less probability with the AF relaying when SNR is high; however all error events are equally probable with the DF relaying. Further, the authors  in \cite{BER_3_IslamIET2016} studied the bit-error-rate of the same channel with AF relaying assuming that the channel estimation is imperfect. It was shown that the average bit-error-rate is linearly increased with channel estimation error.

In \cite{BER_4_IslamSSP2014}, the authors considered the same channel model as in \cite{BER_2_IslamIET2013,BER_3_IslamIET2016} with AF relaying. To reduce interference, the authors proposed a user pairing scheme to improve the bit-error-rate. In specific, the user pairing scheme was proposed to find a user according to its average channel gain to form pairs.

In \cite{BER_5_LouieTWC2010}, the authors compared the performance of the two-time-slot, three-time-slot, and four-time-slot TWRCs from the perspective of outage probability, sum-rate and total bit-error-rate. The obtained results showed that the two-time-slot PLNC scheme provides a larger sum-rate, but worse bit-error-rate than the four-time-slot scheme. The three-time-slot PLNC scheme achieves a good compromise between them. The authors in \cite{BER_6_ParkVTC2011} also studied the bit-error-rate performance of the TWRC but they focused on the two-time-slot PLNC scheme.


In \cite{BER_7_GuanTWC2011}, the authors studied the bit-error-rate of the TWRC with multiple relay nodes when differential binary phase-shift keying modulation was employed. To minimize the average bit-error-rate, the authors also investigated the power allocation by minimizing the asymptotic bit-error-rate at high SNR. By developing upper and lower bounds on the bit-error-rate, the authors demonstrated that the diversity order is exactly $\lceil \frac{K}{2}\rceil$ where $K$ is the number relay nodes.

\subsubsection{Outage probability}

In \cite{BER_1_Hasan2015}, the authors considered the mRC with AF relay. By using the coded slotted ALOHA scheme, the authors proposed an uncoordinated communications and used the iterative demapping algorithm to improve the success rate probability at the user ends. The authors evaluated the bit-error-rate performance of the proposed schemes and showed that the proposed scheme works well even at a relatively low SNR.

In \cite{Outage_1_WangCL2012}, the authors investigated the outage performance of the mRC with a compute-and-forward relay strategy. The outage probability of the compute-and-forward mRC and the non-network-coding mRC were derived and compared. It was shown that the compute-and-forward relay strategy can significantly outperform the non-network-coding mRC.

In \cite{Outage_2_AmarasuriyaTCOMTCOM2013}, the authors studied the performance of the MIMO AF mRC with two zero-forcing transmission schemes, where one scheme utilizes the zero-forcing in each pair and the other uses the zero-forcing in an non-pairwise manner.
The pairwise zero-forcing transmission refers to a mRC where both the MAC phase and the BC phase require $K-1$ time slots with $K$ being the number of user pairs. In the MAC phase, the two users in a pair use zero-forcing precoding to cancel the uplink interference. Then the relay forwards the network-coded signal in the BC phase. Different from the pairwise zero-forcing scheme, the non-pairwise zero-forcing scheme based mRC only requires one time slot in the MAC phase and $K-1$ time slots in the downlink phase. The analytical performance including the outage probability, the achievable sum rate, and the diversity-multiplexing trade-off were obtained in closed forms.

In \cite{Outage_3_ZhaoTWC2011}, the authors studied the mRC with single-antenna users and a multi-antenna relay. A special relay precoding matrix was designed such that the messages in a pair can be aligned. The analytic results of both the ergodic sum rate and the outage probability were derived. The authors in \cite{Outage_4_AmarasuriyaTWC2012} also studied the pairwise AF mRC. A new end-to-end SNR expression at each user was first obtained. Then the approximately analytical expressions of the probability density function and the moment generating function of the channel were obtained to get insightful statistical properties of the considered channel.

In \cite{Outage_9_WangCL2011, Outage_13_Wang2011}, the authors studied the outage probability of the multi-hop TWRC with analog network coding (ANC). The authors derived the closed-form outage probability and demonstrated that the ANC scheme is superior to the traditional AF scheme.

In \cite{Outage_10_Kong2012}, the authors considered a two-way relay channel where space-time coding is utilized to aid the PLNC. The outage probability was analyzed with DF relaying under Rayleigh flat fading.

In \cite{Outage_12_ZhangTWC2016}, the authors studied the TWRC and one-way relay channel with residual self-interference assuming that the relay operates in the full-duplex mode. It was revealed that when SNR is medium or high, the TWRC achieves higher average rate than one-way relay channel, but suffers from certain outage performance loss. In addition, the DF relaying achieves better outage performance than the AF relaying, but suffering from certain rate loss when SNR is high.

In \cite{Outage_14_WangICC2012}, the authors studied the outage performance of the AF TWRC with mutli-antenna relay. Two bidirectional protocols, namely, two-time-slot multiple access broadcast (MABC) protocol and three-time-slot time division broadcast (TDBC) protocol, were analyzed. Two scenarios where the relay node uses fixed gain and zero-forcing (ZF) precoding matrix were discussed.


\subsubsection{Diversity and multiplexing tradeoff}

In \cite{DMT_2_LinTVT2013}, the authors analyzed the outage probability and the finite-SNR diversity multiplexing tradeoff for the TWRC with both AF and DF relay strategies. The authors derived outage probability lower bounds for single- and multi-antenna relay.
Based on the tight lower bounds, the authors further derived estimates of the finite DMTs. The obtained results showed that AF always outperforms DF in terms of finite DMT in all SNR regimes.

In \cite{DMT_5_WangTCOM2011}, the authors analyzed the diversity-multiplexing tradeoff for the DF three-time slots two-way relay channel. The exact outage probability was derived. Then by computing the limiting slope of outage probability versus SNR as the SNR approaching infinity, the authors obtained the DMT. The obtained results showed that network coding can assist the relay to improve the multiplexing and diversity gain.

In \cite{DMT_6_YadavCL2014}, the authors analyzed the DMT of the TWRC when channel estimation errors present. With the derived outage probability expression, the finite-SNR DMT was obtained. The asymptotic and limiting behavior of the DMT was also analyzed to highlight the effect of imperfect channel estimation.

In \cite{Rate_SISO_16_VazeTIT2011}, the authors first proposed an iterative algorithm to maximize the achievable rate of three-time-slot TWRC where each node may have one or multiple antennas. Capacity scaling law with an increasing number of relays was
established based on the capacity lower and upper bounds. Then, the authors proved that the optimal DMT is able to be achieved with the compress-and-forward strategy.

\section{Communication-Theoretic Studies in Multi-way Relay Channel}\label{communication_theoretic}

In this section, we discuss the communication theoretic studies of the mRC. In specific, we focus on the channel estimation, power allocation, precoding/beamforming design and relay/antenna selection.

\subsection{Channel Estimation}
CSI is of critical importance for the power allocation, precoding/beamforming design and relay/antenna selection.
%
As two-hop transmission is employed in the mRC, the CSI of both the two hops needs to be estimated.
The channel estimation in a point-to-point wireless communication system is usually carried out as follows: The transmitter sends a period of known signals-called pilot signals or training sequence; upon receiving the pilot signals, the receiver estimates the channel.
In the mRC, if the relay has a capability of conducting channel estimation procedure, the two-hop channel estimation can decoupled into a simple one-hop channel estimation problem. That is, in the MAC phase, all users send training sequences to the relay simultaneously, and the relay performs the channel estimation process after receiving the training sequences to estimate the channel coefficients from the users to the relay ; in the BC phase, the relay sequences are transmitted from the relay, and each user received the training sequences to estimate the channel coefficients from the relay to the user.
%

On the other hand, if the relay is not able to conduct the channel estimation process, the channel estimation procedures should be performed at the user ends. As the channel coefficients of two hops needs to be estimated at the user ends simultaneously, the channel estimation problem is more challenge. Judicious design of the training sequences is required to estimate the channel of the two hops.

We next discuss the existing channel estimation strategies for the mRC in the literature.
The channel estimations for single-antenna single-carrier TWRC was investigated in \cite{Channel_estimation_GaoTCOM2009, Channel_estimation_XieTWC2014}. In \cite{Channel_estimation_GaoTCOM2009},
the authors considered the scenario where the channels exhibit the time reciprocal property, i.e., the channel from a user to the relay is identical to the channel from the relay to this user. Under this setup, the involved four unknown channel coefficients reduces to two unknown ones.
Instead of estimating two individual channel coefficients, the authors proposed to estimate the cascade channel coefficients of two hops.
The channel coefficients were estimated with criteria of maximum-likelihood (ML) and maximum average effective SNR. The training sequences were further optimized to minimize the Cramer-Rao lower bound (CRLB) and maximize the average effective SNR. It was shown that the optimal training sequences at two users exhibit an orthogonal property. In \cite{Channel_estimation_XieTWC2014}, the Bayesian estimation method was applied to estimate the cascaded user-relay-user channel. By assuming with a priori knowledge of channel distribution, the cascaded channel coefficients and the amplitude of individual channels were estimated with maximum a posteriori (MAP) based estimation schemes.

In \cite{Channel_estimation_JiangTWC2010},  the authors developed a channel estimation prototype for the AF TWRC. By allowing the relay to first estimate the channel coefficients using the ML channel estimation, power allocation was performed to maximize the average effective SNR for achieving an efficient data detection.

In \cite{Channel_estimation_WangGTWC2011}, the authors studied the channel estimation for the AF TWRC with time-selective fading. A complex-exponential basis expansion model was utilized to characterize the time-varying channel coefficients, so as to reduce the free variable number needing to be estimated in the channel acquisition.

In \cite{Channel_estimation_GaoTSP2009}, the authors studied the channel estimation for the multi-carrier TWRC, where the orthogonal frequency division multiplexing (OFDM) modulation is applied to assist the communication. Only the two users are allowed to transmit pilot sequences. The authors developed block based and pilot-tone based approaches to estimate the cascaded user-relay-user channels and the individual channels, respectively. The estimation ambiguity analysis showed that, provided that the training sequences are sufficient long, only the sign ambiguity occurs, which does not affect the final data decoding.

%

In \cite{Channel_estimation_ZhangSTSP2012}, the authors considered to jointly estimate the individual channels of the OFDM TWRC. The authors characterized the bayesian CRLB for the individual channel estimation, and the optimal training structure to minimize the bayesian CRLB was found.

The authors in \cite{Channel_estimation_Rui2015TWC} investigated the channel estimation for the MIMO single-carrier TWRC by taking the impact of the neighboring users' interference into account. In addition, the estimations were performed in both uplink and downlink phases such that the individual MIMO channel can be estimated. The linear minimum mean square error (LMMSE) estimators were developed for each phase. The effect of the training length on the the estimation performance were discussed.

In \cite{Channel_estimation_PhamTCOM2010}, the authors proposed to estimate the composite channels of the MIMO TWRC. Two kinds of composite channels, i.e., user one-relay-user one and user one-relay-user two, were considered. 
The least-square (LS) principle was applied to estimate the channel coefficients.

In \cite{Channel_estimation_KimCL2013}, the authors designed a training signal to minimize the total MSE of the correlated MIMO AF TWRC.
%
The training signals' optimal structure was presented.

The channel estimation of the MIMO TWRC was extended to the multi-carrier case in \cite{Channel_estimation_KangTCOM2017}. Instead of estimating individual channels, the convolution of two individual MIMO channels was estimated in \cite{Channel_estimation_KangTCOM2017} by using the self-interfering link and the information-bearing link. With the LMMSE criterion, the training sequences were optimized to minimize the total MSE at two users subject to user and relay power constraints. The training sequences' optimal structure was analyzed and utilized to reduce the optimization complexity.

\subsection{Power Allocation}

The transmission powers of the nodes in the mRC need to be allocated appropriately to enhance the quality-of-service (QoS) of the system. The QoS can be measured by a number of metrics, such as achievable sum-rate, outage probability, SNR, bit-error-rate, etc. Various optimization techniques have been developed to tackle the power allocation problem in the mRC, as detailed below.


The authors in \cite{Power_alloca_ge_1_Shin2009} studied the power allocation in the TWRC employing the PLNC protocol. The power was optimized with an aim to maximize the achievable sum rate of the DF-based PLNC protocol subject to the sum power constraint.
The authors in \cite{Power_alloca_ge_2_Yuan2011} optimized the power aiming to maximize the achievable sum rate of
the TWRC with AF-based PLNC. The authors in \cite{Power_alloca_ge_3_Pischella2013} investigated the power allocation at both the users and the relay to maximize the sum rate, considering the sum power and the fairness constraints with both AF and DF relay strategies. The authors in \cite{Power_alloca_ge_4_Wilson2009} obtained a capacity upper bound of the TWRC under the sum power constraint.
The authors in \cite{Power_alloca_ge_5_Peh2008}  derived the bit-error-rate performance of the
TWRC under the Rayleigh fading environment. They also obtained the optimal power allocation to
minimize the instantaneous bit-error-rate of the PLNC scheme.
The authors in \cite{Power_alloca_ge_6_Kim2015} proposed approximation methods for the bit-error-rate of the MAC phase in
fast fading channels, and the transmit power allocation
strategy for each source node was optimized in order to minimize the derived
bit-error-rate.
The authors in \cite{Power_alloca_ge_7_Cetin2012} studied the power allocation in the AF TWRC. Closed-form expressions for
the optimal powers were derived by solving the total power minimization problem.

The frequency selectivity results in inter-symbol interference (ISI) at receivers. A common approach to combat the ISI in frequency selectivity TWRCs is the OFDM technology which transforms the end-to-end channel into a set of parallel sub-channels.
The authors in \cite{Power_Allocation_OFDM_1_Dong2010} considered the power allocation problem for the OFDM TWRC with AF relaying strategy subject to the total network power constraint. They obtained the optimal power allocation across the subcarriers of all the three nodes to maximize the achievable sum rate in the network. Moreover, the authors demonstrated that the resulting solution is a combination of two power allocation strategies, i.e., water-filling across subcarriers and SNR-balancing between the two user nodes. The authors in \cite{Power_Allocation_OFDM_2_Ho2008} considered subcarrier permutation at the relay as well as power allocation across the subcarriers with respect to the individual power constraints of each node. The authors in \cite{Power_Allocation_OFDM_3_Liu2017,Power_Allocation_OFDM_4_Cai2013,Power_Allocation_OFDM_5_Alihemmati2015} considered a TWRC model with $K$ relays, where the authors in \cite{Power_Allocation_OFDM_3_Liu2017} and \cite{Power_Allocation_OFDM_4_Cai2013} considered selecting the best relay among several relays, and the authors in \cite{Power_Allocation_OFDM_5_Alihemmati2015} considered a joint power allocation and network beamforming design. Furthermore, the authors in \cite{Power_Allocation_OFDM_4_Cai2013} considered the problem of joint relay selection, subcarrier pairing, power allocation, and bit loading.

The authors in \cite{Power_Allocation_OFDM_6_Sidhu2011,Power_Allocation_OFDM_7_Zhang2012, Power_Allocation_OFDM_8_Shin2010} considered the power allocation problem in the multi-user TWRC. Specifically, the model studied in \cite{Power_Allocation_OFDM_6_Sidhu2011} consists of multiple pre-assigned pairs of mobile users and one fixed relay station. The optimization target was to find the best subcarrier allocation to each user, subcarrier pairing at the relay, as well as the power allocation at all nodes.  The authors in \cite{Power_Allocation_OFDM_7_Zhang2012} considered a single-cell multi-carrier TWRC model consisting of one common BS, multiple mobile stations and multiple relays. Each relay operates in a half-duplex mode and forwards the bi-directional traffic using the AF protocol. The relay-power allocation, relay selection, and subcarrier assignment were jointly optimized to enhance the performance. Taking the user fairness as constraints, the authors in \cite{Power_Allocation_OFDM_8_Shin2010} allocated the subcarriers to each user-pair and relay to maximize the achievable sum rate.

Traditional OFDM based TWRCs only considered a relay strategy in each subcarrier and treated all subcarriers as independent channels. This scheme results in significant performance losses, especially when the channel experiences fading. In \cite{Power_Allocation_OFDM_9_Chen2017}, the authors developed a generalized subcarrier pairing strategy such that a user-pair can occupy multiple subcarriers during the uplink phase and downlink phase transmissions. Resource allocation was then conducted to improve the spectrum efficiency and achieve additional multi-user diversity gain.

\subsection{Precoding and Beamforming}

In this subsection, we survey recent process on beamforming and precoding strategies in the mRC. In general, the studies of the beamforming/precoding in the mRC can be classified by the target performance metrics, such as mean-square error, achievable rate, max-min fairness, etc.

In \cite{Beam_2_WangJSAC2013}, the authors considered the cellular TWRC, where a multi-antenna BS exchanges messages with a couple of users with the assistance of multiple relays. The transceiver at BS, relays and users were first jointly optimized to minimize the weighted sum MSE. To extending the proposed algorithm to sum rate maximization problem, the authors established the equivalence between the sum rate maximization and the weighted sum MSE minimization.
The same cellular TWRC model was considered in \cite{Beam_4_WangTWC2013}. In specific, the authors in \cite{Beam_4_WangTWC2013} showed that the precoding design at BS can be optimally solved using second order cone programming (SOCP) and the relay beamforming matrix can be solved by semidefinite programming (SDP).

In \cite{Beam_6_JoungTSP2010}, the authors studied the multipair TWRC where multiple single-antenna users exchange messages via a multi-antenna relay. The relay beamformer was optimized based on ZF and minimum mean-square-error (MMSE) criteria. The same channel model was also considered in \cite{Beam_5_TaoTSP2012} where the relay beamforming matrix was found by using SDP with an aim to achieve performance fairness among users.

To achieve near-optimal performance, the authors in \cite{Beam_1_FangTSP2013} proposed a Dinkelbach-type algorithm to optimize the relay beamforming matrix. By taking the max-min SINR solution as a basic component, the authors showed that the proposed Dinkelbach-type beamforming design algorithm can be extended to solve weighted sum-rate maximization, weighted sum MSE minimization, and average bit-error-rate minimization problems. The key procedure for the extension was to transform the beamforming design problem with various criteria to a monotonic program, which can be efficiently solved using the polyblock outer approximation algorithm.


In \cite{Beam_8_DingTWC2011}, the authors considered the multipair TWRC model as \cite{Beam_6_JoungTSP2010, Beam_1_FangTSP2013}. The authors showed that when the sources have less antennas than the relay, the signal alignment can be used to design beamforming matrices. The authors proved that the proposed scheme achieves a higher multiplexing gain than the time sharing approach.

In \cite{Beam_13_WangTIT2013}, the authors considered the multipair TWRC with an arbitrary unicast data exchange model.
Similarly to \cite{Beam_2_WangJSAC2013, Beam_4_WangTWC2013}, the authors developed an iterative algorithm to jointly optimize the relay beamforming matrix and users' receiving matrices to minimize the weighted MSE or maximize the weighted sum rate.

In \cite{Beam_3_WangTSP2012}, the joint design of the source, relay and the user processing matrices was investigated to minimize the total MSE. The authors proposed an iterative algorithm to find a locally optimal solution by decomposing the original joint optimization problem into three subproblems. It was shown that the source precoders can be optimally solved via quadratically constrained quadratic program (QCQP) and the optimal relay precoder can be obtained using the Karush-Kuhn-Tucker (KKT) conditions.

The same channel model was also considered in \cite{Beam_10_XuTWC2011}. In specific, the authors in \cite{Beam_10_XuTWC2011} aimed to maximize the weighted sum rate. It was revealed that when the relay precoder is given, the source precoders can be found with the generalized water filling algorithm. Alternatively, the relay precoder can be solved using the hybrid gradient algorithm together with the Newton's search. The nonlinear precoding design was studied in \cite{TVT_2012_Wang} assuming that the MMSE decision feedback equalizers are used at two user ends, while linear precoding is employed at the source and relay nodes. With given relay precoder, the authors proved the convexity of the nonlienar source design problem, and developed an iterative algorithm to obtain the optimal solution. Then with the given source precoder, the optimally analytical solution of the relay precoder was found with the KKT conditions.

In \cite{Beam_9_ZhangJSAC2009}, the authors studied the relay beamforming matrix optimization for the TWRC assuming users with single antenna. The weighted sum rate was maximized by converting the original non-convex problem into a relaxed SDP problem, based on which the optimal relay beamformer was finally obtained.

\subsection{Relay and Antenna Selection}

For the mRC with multiple relays or multi-antenna relay, relay selection or antenna selection have been studied to improve the end-to-end performance of the system.

The authors in \cite{BER_10_ZhouTCOM2010} studied a DF multi-relay TWRC network. A opportunistic strategy was proposed to select the ``best relay" according to the max-min criterion. The PLNC was performed at the chosen relay to achieve better performance.

The authors in \cite{RelaySelection_7_Atapattu2013} studied a AF multi-relay TWRC. A single-relay selection scheme was first developed to maximize the the worse SNR of two users. Then a more complicated multiple relay selection scheme was proposed to achieve a better performance. A couple of performance metrics, including the block error rate, the diversity order, the outage probability, and the sum-rate, were analyzed based on the derived cumulative distribution function of the worse SNR at the two users.

The authors in \cite{BER_12_LiWT2014} investigated the relay antenna selection for an asymmetric mRC with analogue network coding strategy. The authors showed that by using the orthogonal projection technique, the minimal required relay antenna number can be smaller than the total data stream number. A optimal relay antenna subset selection scheme was developed based on the paired throughput max-min criterion and the authors proved that full diversity gain is achieved by using the the proposed scheme.

The authors in \cite{RelaySelection_5_Sheikh2016} studied the multi-antenna and multi-relay TWRC with MABC protocol. A relay selection with ML detector used at the relay was proposed and the corresponding bit-error-rate performance was analyzed. The authors showed that the proposed relay selection algorithm achieves full diversity order.

The authors in \cite{RelaySelection_1_CaiTVT2018} investigated a two-relay TWRC with differential chaos shift keying (DCSK). With the DF relaying strategy, the authors analyzed the performance metrics of bit-error-rate, diversity order, and throughput. It was showed that two-relay TWRC DCSK system could achieve lower bit-error-rate, but degrade the throughput meanwhile as compared with traditional one-relay TWRC DCSK system. To further improve the performance, the authors developed a non-CSI required relay selection criterion and the performance of the max-sum, max-product and max-min relay selection schemes were analyzed.

The authors in \cite{RelaySelection_2_DoCL2013} considered a multi-relay TWRC adopting the DF relaying strategy. Relay selection and power allocation were jointly optimized to improve the symbol error probability performance. The performance of the benchmark scheme considering only relay selection while without power allocation was also analyzed accordingly.

The authors in \cite{RelaySelection_8_Wang2013} investigated the joint power allocation and relay selection for the AF TWRC. The asymptotic symbol error probability was characterized with the second-order channel statistics. The derived asymptotic symbol error probability wwas further used as a criterion to optimize the power allocation and the relay selection. To achieve full diversity
order, the authors proposed an alternative power allocation and relay
selection method by simply replacing the statistical channel
information with an instantaneous one. The corresponding diversity order was acquired by using
the lower and upper bounds of the asymptotic outage probability.

The authors in \cite{BER_9_ZainaldinTWC2016} studied the relay selection for relay assisted LTE-advanced networks. The considered network assumes that the information exchange between user equipment and eNodeB is conducted using three possible transmission schemes, i.e., direct transmission, cooperative relaying, or a combination of network coding and cooperative relay. With an aim to maximize the produce of backlog and rate, the joint combinatorial optimization problem associated with the relay selection, and the bidirectional
transmission scheme selection was formulated. A graph-based framework and a hybrid ant colony optimization algorithm were developed to solve the established combinatorial optimization problem.

\section{Open Issues and Future Extensions}\label{open_issue}
In this section, we discuss several possible future directions in this field, including analyzing synchronization in the mRC, the DoF of the mRC with delayed CSI, the channel-unaware mRC, the multiple-relay/multiple-hop mRC, applications in MEC and Fog-RAN, and applications in the IRS.

\subsection{Synchronization in the mRC}

The above discussion on the mRC assumes perfect synchronization between nodes. Under this assumption, their packets should reach the relay with aligned the packet boundary and symbol boundary. To achieve this alignment, different levels' synchronization is required to be feasible, including MAC-layer packer alignment, symbol level alignment, and RF carrier frequency synchronization. In general, symbol alignment is more challenging as it operates at a finer time scale than packet alignment. Imperfect RF carrier frequency synchronization results in a phase offset at the relay nodes, which may degrade the system performance. synchronization studies of two-way relay node can be found in \cite{Synchronization_Zhang}, but extending to the multiuser mRC is an interesting future extension.

\subsection{Delayed CSI}\label{Delayed_CSI}

In the mRC, when performing self-interference cancelation, the end nodes usually need to know the CSI of the whole network. To this end, we usually assume that there is a dedicated central processing unit, which could collect the entire network's CSI. In general, obtaining CSI using the dedicated central processing unit can incurs delay for that it is required for the channel estimation and the estimated CSI feedback. Effect of CSI delay on DoF were studied in \cite{Delay_LiCL2014, Delay_LiCL2016, Delay_SPLWang}. In addition, the work \cite{Delay_SPLWang} illustrated that the interference neutralization is effective in compensating the performance loss introduced by delay CSI. However, for a general case of delayed CSI, the DoF study remains unsolved. 

\subsection{Channel-Unaware mRC}

It is noted that the key technique to achieve the performance gain in the mRC is the self-interference cancelation. As mentioned before, self-interference cancelation requires the perfect CSI of the entire network. In subsection \ref{Delayed_CSI}, we have discussed to use delayed CSI to achieve the gain. In an extreme scenario, the CSI may be totally unavailable. In this case, how to cancel the self-interference cancelation is an interesting direction. One promising solution is to apply the differential scheme like in traditional differential space-time coding \cite{Bameri_TWC_2019}, i.e., each use ends use the received signal at previous time slot to help cancel the self-interference in the current time slot. For this scheme, the performance degradation brought by unknown CSI should be analyzed.

\subsection{Multi-Relay/Multi-Hop mRC}


In the mRC, if the processing capability of a single relay is limited, considered using multiple relays to assist the transmission may be useful \cite{Gao_access_2019, Wang_IOT_2020}. Further, when the distance between the users are too long, the messages delivery may need multiple hops, while not just two hops. Therefore, extending the mRC to multi-relay and multi-hop scenarios is critically necessary to widen the application of the mRC \cite{Chang_Access_2019, Chen_VTC_2019}. Undoubtedly, extending corresponding information theoretic studies and communication theoretic studies to multi-relay and multi-hop scenarios is the first step to achieve this goal.

\subsection{Applications in MEC and Fog-RAN}

Mobile edge computing \cite{Zhou_Proceeding_2019, Park_Proceedings_2019, Elbamby_Proceedings_2019, Li_TWC_2020} and fog radio access network \cite{Abdel_Access_2019, Chiu_TSC_2019, Sun_TVT_2019, Rui_TVT_2019} have been considered as key techniques in future cellular communications. The key feature of the MEC and the F-RAN is to introduce the caching ability and computation ability in the edge networks. This feature brings a nature advantage to combine with the mRC. In both the MEC and the Fog-RAN, the simple consideration is that the BS is capable of performing caching and computing. In this case, multiple mobile users exchanging messages via the BS can be modeled as a mRC where the BS performs as a relay node. Furthermore, the cached messages at the BS can help to generate the physical-layer networked signals, which may provide more degrees-of-freedom to use the computation ability to enhance the network performance.

\subsection{Applications in IRS}

Intelligent reflecting surface defines a class of artificial passive radio structures that can consume very low power to reflect incident radio frequency waves to specified directions \cite{IRS_liangJCIN2019,  IRS_arxivBasar2019, IRS_ComMLiaskos2018, IRS_RenzoEURASIP2019, IRS_qqwuCOMM2020, IRS_Yuanarxiv2020, IRS_ErtugrulBasararxiv2019, IRS_HuTSP2018, IRS_WuTC2019, IRS_YuanWCL2020, IRS_YuanWCL2020_2, Mishra_ICASSP_2019}. IRS is also called reconfigurable smart surface or large-scale smart surface,
which is actually a planar array consisting of a large number of passive reconfigurable elements. Each of the elements is able to induce a certain phase shift independently on the incident signals, thus collaboratively change the reflected signal propagations.
Although the function of IRS is similar to the full-duplex AF relay, the key difference between them is that the radio frequency reflection on the IRS is passive, nearly introduces no additional thermal noise and consumes no power. Hence, IRS has been identified as a key component to construct the smart wireless communication environment and improve the transmission performance.
%
%
The IRS can also be treated as a passive relay node to assist the transmissions among multiple users. The formed network can be modeled as an mRC. The key difference from the traditional relay is that the passive IRS only adjusts the phase of incident signals. Hence, the IRS assisted mRC brings many new problems for both information-theoretic and communication-theoretic studies which deserves new research effort.

\section{Conclusion}\label{conclusion}

In this article, we provided an overview of the studies on the mRC. We began with several classic application scenarios which can be modeled by the mRC. Then, we established a general system model for the mRC with arbitrary data exchange patterns. After that, we surveyed the state-of-the-art of the mRC from aspects of the information-theoretic and communication-theoretic studies. Specifically, key results of achievable-rate/capacity analysis, degrees-of-freedom analysis, channel estimation, power allocation, precoding/beamforming design, relay and antenna selection were summarized and discussed. The related research challenges and future extension were finally presented.

In summary, in spite of the abundant existing studies on the mRC, the fundamental limits of the mRC with a general data exchange model is still far from being well understood. This article attempts to briefly discuss the current progress of related technologies. We hope that our discussion and exploration may open a new avenue for the development of the mRC.

\bibliographystyle{IEEEtran}
\bibliography{IEEEabrv,ref1}

\end{document}